\documentclass[%
 amsmath,amssymb,
 reprint,
 prx,
 longbibliography,
floatfix,
nofootinbib,
superscriptaddress,
]{revtex4-2}
\usepackage{graphicx}
\usepackage{dcolumn}
\usepackage{bm}
\usepackage{mathrsfs}
\usepackage{amsmath}
\usepackage{enumitem} 
\usepackage{cancel,siunitx}
\let\svqty\qty
\usepackage{physics}
\let\qty\svqty
\usepackage[dvipsnames]{xcolor}
\usepackage{blkarray}
\usepackage{multirow,bigdelim}
\usepackage{soul}

\newcommand{\abinit}[0]{\textit{ab initio }}

\newcommand{\I}[0]{\int d\mathbf{r}}
\newcommand{\pbar}[0]{\overline{p}}
\newcommand{\qbar}[0]{\overline{q}}
\newcommand{\rbar}[0]{\overline{r}}
\newcommand{\sbar}[0]{\overline{s}}

\newcommand{\kbar}[0]{\overline{k}}

\makeatletter
\def\maketitle{
\@author@finish
\title@column\titleblock@produce
\suppressfloats[t]}
\makeatother

\begin{document}

\preprint{APS/POL-DHF}

\title{A comprehensive theory for relativistic polaritonic chemistry: a four components \abinit treatment of molecular systems coupled to quantum fields}

\author{Guillaume Thiam}
\email{guillaume.thiam@unipg.it}
\author{Riccardo Rossi}%
\affiliation{Dipartimento di Chimica, Biologia e Biotecnologie, Universit\`a degli Studi di Perugia, Via Elce di Sotto, 8,06123, Perugia, Italy}

\author{Henrik Koch}
\affiliation{Department of Chemistry, Norwegian University of Science and Technology, 7491 Trondheim, Norway}

\author{Leonardo Belpassi}
\affiliation{Istituto di Scienze e Tecnologie Chimiche “Giulio Natta” del CNR (CNR-SCITEC), Via Elce di Sotto, 8, 06123 Perugia, Italy}

\author{Enrico Ronca}
\email{enrico.ronca@unipg.it}
\affiliation{Dipartimento di Chimica, Biologia e Biotecnologie, Universit\`a degli Studi di Perugia, Via Elce di Sotto, 8,06123, Perugia, Italy}


\begin{abstract}
We present an \abinit approach to study molecules containing heavy atoms strongly interacting with quantum fields in optical devices. The theory has been derived from the relativistic quantum electrodynamics (QED) introducing the approximations needed to provide a formalism suitable for relativistic quantum chemistry. This framework represents the ideal starting point to extend the main quantum chemistry methods to relativistic polaritonic.
In this article, we present the Polaritonic Dirac Hartree Fock (Pol-DHF) based on this theory. Pol-DHF approach allows for the simulation of field induced effects on the ground and excited state properties of heavy transition metals molecular complexes. The method is able to include not only the effects of the photons, but can in principle be extended also to include explicit interactions with positrons.
Application of Pol-DHF to three metal hydrides shows that the magnitude of both polaritonic and relativistic effects can be comparable when relativistic effects are getting more important. Due to an accurate description of spin-orbit coupling, the method is able to reproduce polaritonic effects happening at the crossing between singlet and triplet potential energy surfaces.  
\end{abstract}

\maketitle


\section{\label{sec:level1} Introduction}
The use of light as a new tool to control and manipulate non-invasively the properties of molecules and materials is opening, in recent years, a new field of research at the border between physics, chemistry and material science~\cite{RydCavHaroche1985, LiuNatureCavityExp2015, EbbesenACR2016, ThomasAngewChem2016, lather2019cavity}.
\begin{figure}[ht!]
    \includegraphics[width=0.85\linewidth]{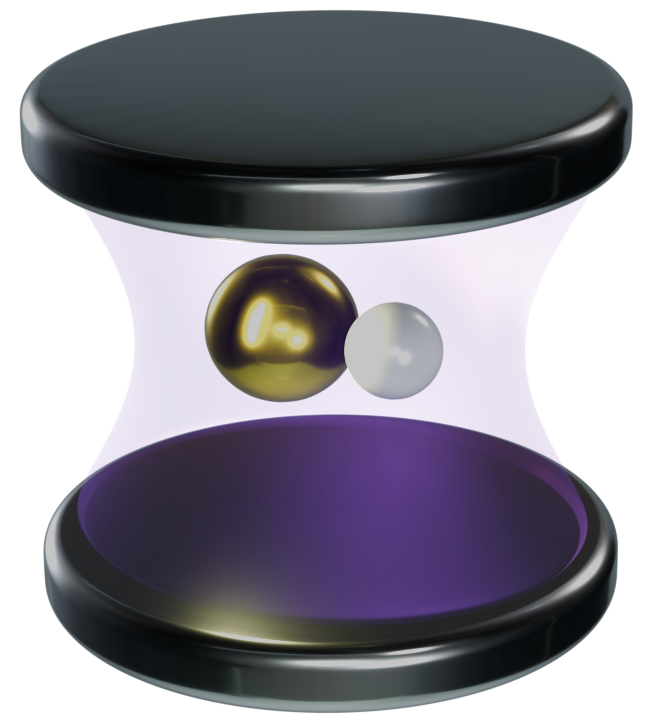}\hfill
    \caption{Schematic representation of a Fabry-Pérot cavity containing a gold-complex.}
    \label{fig:cavity}
\end{figure}
When matter strongly couples to photons, new hybrid states (polaritons), having partial light and partial matter character, are formed. The strong coupling condition is usually reached inside properly designed optical devices. The simplest example is the Fabry-Pérot cavity~\cite{Optical_Cavity_1992_Sci} (Fig.~\ref{fig:cavity}), made of two highly reflective planar mirrors, that confine the photons, leading to a significant enhancement of the light-matter coupling.
The cavity frequency, and therefore, the polaritons properties can be controlled changing the geometry and the materials in the device. Several manifestations of polaritonic effects on different physical properties such as absorption spectra, photochemical reaction rates, and conductivity have already been observed in the experiments~\cite{LiuNatureCavityExp2015,fregoni2018manipulating, EbbesenACR2016,lather2019cavity,herrera2020photochemistry,delpo2021polariton,sentef2018cavity}. 
What is probably the most well-known demonstration of such effects was obtained in the experiments performed by Ebbesen's group.  In particular, they evidenced that strong coupling to molecular vibrations can be used to catalyze, slow down or even induce selectivity in chemical reactions~\cite{thomas2019tilting,lather2019cavity,ahn2023modification}. 
These observations opened a new field that is now known as polaritonic chemistry~\cite{Polaritonic_org_mole_2018}. \\

In these experiments, the photonic states are usually coupled either to electronic or vibrational states of the molecular systems. However, the electromagnetic nature of the field also allows for modifications of their magnetic properties if one exploits the coupling to spin-states. In this way, a fine control of the magnetizability and aromatic properties of molecules~\cite{barlini2024theorymagneticpropertiesqed}, of  spin qubits~\cite{PetersonNaturecQED2012,AffronteCircuitQEDSciRep2017} and of spin phases of materials~\cite{HubenerMatQuantumTech2024, PhysRevXphasetransitionQEDLattice2017} can be obtained.   
However, reaching the strong coupling condition in this frequency range is impossible using a simple Fabry-Pérot cavity that, in this case, would require a spacing of centimeters between the mirrors. The problem can be circumvented using planar superconducting devices commonly used in Circuit-QED experiments~\cite{CircuitQED_Review2021}. Using similar devices, Affronte's group has been able to manipulate the spin properties of a Single-Molecule Magnet~\cite{AffronteCircuitQEDSciRep2017, AffronteAdvPhysX2018}. Such an accomplishment unlocked many potential applications in spin qubits based quantum computation. 

In spite of the many improvements in the fabrication of more effective optical devices and the impressive accuracy reached~\cite{JCPWeichman2023, jacsWeichman2023} by polaritonic chemistry experiments, many fundamental aspects still remain to be understood. In this context, theory represents a fundamental tool to gain insights on the underlying physics of these processes. In recent years, many \abinit methods, able to treat electron-electron and electron-photon correlation at different level of accuracy, have been developed~\cite{Flick_PRA_QED_DFT_2014, phys_rev_x_Ronca_2020,DePrince_JCP_2021_QED_CC_IP_EA, Flick_JPCL_QED_CC_2021,nature_riso_2022,HauglandJCP2021, alessandroqedcas_2025}. However, they have mainly been applied to investigate cavity induced effects on the electronic and vibrational degrees of freedom of molecular systems. Only very recently, Barlini et al. proposed the first Hartree-Fock based approach to study photon induced effects on the electronic and nuclear magnetic properties of a molecular system~\cite{barlini2024theorymagneticpropertiesqed}.  

Spin orbit coupling and to some extent magnetic properties can be seen as a manifestation of relativistic effects in molecules. 
Therefore, an accurate investigation of these properties requires the inclusion of relativistic effects at different level of accuracy. 
This becomes particularly crucial when, like in Affronte's experiments~\cite{AffronteCircuitQEDSciRep2017,AffronteAdvPhysX2018, BonizzoniMem2020, Bonizzoni2021, Bonizzoni2022, Bonizzoni2023, Bonizzoni2024} Single-Molecule Magnets containing lanthanides atoms need to be used to have sufficiently long lasting magnetizations.

Moreover, as already discussed in Refs~\cite{HauglandJCP2021, haugland2023Arkiv} for Van-der Waals interactions, the quantum electromagnetic field can sometimes enhance small effects that are usually negligible in the absence of photons. 
From these considerations rises the need of a consistent relativistic quantum electrodynamical \abinit theory able to include all the necessary effects. 
Formulating such a general approach is one of the goals of this paper. 
Similar intent was already presented by Ruggenthaler et al.~\cite{Flick_PRA_QED_DFT_2014}
in a Density Functional Theory (DFT) framework, but only the non-relativistic limit of the method was actually turned into a usable quantum chemistry implementation. Very recently, Konecny et al.~\cite{RelQEDDFT} proposed a relativistic response theory to investigate electronic excitations of relativistic molecules in optical cavities. Though this last approach is able to capture interesting effects like cavity-induced singlet-triplet interactions, at the moment, the implemented approach does not account for modifications induced by the field on the system's electronic structure. This can be of some importance when investigating molecular properties such as the ionization potential/electron affinity~\cite{DePrince_JCP_2021_QED_CC_IP_EA}.

In this paper, we develop the first wave-function based relativistic \abinit method describing the ground state of molecular systems coupled to photons in optical cavities. We start the development from the QED Lagrangian to then propose a general relativistic formalism. The latter represents the starting point for the development of mean-field but also, in the future, exact 2-components and correlated methodologies able to perform accurate simulation of these complex systems. We believe that this represents a step forward in the study of polaritonic effects that could have repercussions in various areas of science such as Quantum Computing, OLED technologies, biochemistry, etc.

The paper is structured as follows: firstly, a generic derivation of a relativistic QED Hamiltonian theory will be presented, starting from the standard QED Lagrangian. A lot of care has been dedicated to  present the theory in a clear and understandable way even for a non-expert audience. Then, the Coulomb gauge Hamiltonian, after application of the dipole approximation, is used to develop the first Relativistic Polaritonic HF (Pol-DHF) approach. In this context, some space has been dedicated to the strategies used to deal with negative energy states. Lastly, Pol-DHF has been applied to investigate electronic properties of small diatomic molecules containing heavy atoms. We end the paper with conclusions and perspectives.       


\section{Theory}\label{sec:level2}

In this section, we follow the formal derivation of relativistic QED theory usually presented in physics text books\cite{CohenTanoudjiQED, ReiherBook} to develop a Hamiltonian formalism that can be applied to formulate new \abinit methodologies for the simulation of polaritonic molecular systems. This choice has been meant to render the overall discussion accessible to a broad chemistry audience. The proposed methodology is then used  to develop the first Hartree-Fock (HF) based approach for relativistic molecular systems strongly coupled to quantum fields.  
For convenience reasons, Gaussian units\footnote{[cm, g, s]} will be used during the whole derivation unless specified otherwise. \\ 

In order to develop a quantum theory (see Ref.s~\citenum{CohenTanoudjiQED, ReiherBook} and section 1.1 of the Supporting Information (SI) for more details), one generally needs to start with a Lagrangian formalism from which the fields conjugate momenta can be extracted. Our starting point is the Lagrangian density $\mathcal{L}_{\mathrm{QED}} = \mathcal{L}_{\mathrm{Maxwell}} + \mathcal{L}_{\mathrm{Dirac}} + \mathcal{L}_{\mathrm{int}}$  which contains three distinct contributions. One purely related to electromagnetism (originating from the Maxwell Lagrangian), one purely describing non-interacting fermions (originating from the Dirac Lagrangian) and finally a coupling term allowing to describe interacting fermions (see section 1.1 in SI for more details). From this Lagrangian density, we have constructed the QED Hamiltonian performing a Legendre transform:
\begin{equation}\label{eq:ham_density}
    \mathcal{H}_{\mathrm{QED}}=\Pi_\mu \Dot{A}^\mu + \pi\Dot{\Psi}_e-\mathcal{L}_{\mathrm{QED}}
\end{equation}
where $\mu \in \{0, 1, 2, 3 \}$ and the Einstein's summation convention apply and repeated indices are summed. The conjugate momenta are given by:
\begin{equation}\label{eq:conj_momenta}
    \pi=\frac{\partial\mathcal{L}_{\mathrm{QED}}}{\partial\dot{\Psi_e}} \qquad
    \Pi_{\mu} = \frac{\partial\mathcal{L}_{\mathrm{QED}}}{\partial\dot{A^{\mu}}}
\end{equation}  
the latter are not only necessary to build the Hamiltonian density, but also to build (anti-)commutation relationship needed for the quantum theory. The classical Hamiltonian then reads as:
\begin{widetext}
\begin{align}\label{eq:H}
H & = -\frac{1}{8\pi}\int \left[\underbrace{-\phi \nabla^2 \phi}_{\mathbf{E}^{2}_{\text{long}}} \underbrace{-\frac{\mathbf{\dot{A}}\cdot \mathbf{\dot{A}}}{c^2} - (\nabla\times\mathbf{A})\cdot(\nabla\times\mathbf{A})}_{\mathbf{E}_{\mathrm{trans.}}^{2} + \mathbf{B}^{2}}\right] d\mathbf{r} \nonumber \\
& + \int \Psi_{e}^{\dagger}[c\alpha_{i} (-i\hbar \nabla_i - \frac{e}{c} A_{i} - \frac{e}{c} A^{\text{ext}}_{i} ) +\beta m_{e}c^2 )] \Psi_{e} d\mathbf{r}  + \int \phi \rho d\mathbf{r} + \int \phi_{ext} \rho d\mathbf{r}. 
\end{align}   
\end{widetext}
In Eq.~\ref{eq:H} $i \in \{1,2,3\}$.
 $\alpha_{i} = \begin{pmatrix}
\mathbf{0} & \sigma_{i} \\ \sigma_{i} & \mathbf{0}
\end{pmatrix}$ and $\beta = \begin{pmatrix} \mathbf{1} & \mathbf{0} \\ \mathbf{0} & \mathbf{-1} \end{pmatrix}$ are $4\times4$ matrices. \\ $\boldsymbol{\sigma}_i$ are the Pauli matrices:
\begin{equation}
\sigma_{x}= \begin{pmatrix}
0 & 1 \\ 1 & 0
\end{pmatrix}  \hspace{5mm}
\sigma_{y}= \begin{pmatrix}
0 & -i \\ i & 0
\end{pmatrix} \hspace{5mm}
\sigma_{z}= \begin{pmatrix}
1 & 0 \\ 0 & -1    
\end{pmatrix}
\end{equation}
The auxiliary fields $\phi$ and $\mathbf{A}$ are respectively the scalar and vector potentials. The latter are not uniquely defined, and many potentials lead to the same electric and magnetic field. This is referred to as gauge freedom~\cite{CohenTanoudjiQED, Jackson2002Gauge} (see section 1.1 of the SI for more details).
In Eq.~\ref{eq:H} no gauge choice has been applied so far.
For quantum chemistry applications, the Coulomb gauge ($\mathbf{\nabla}\cdot\mathbf{A} = 0 $) is the natural choice. The latter allows splitting the electric field into a longitudinal and a transversal component. This then gives the instantaneous Coulomb interaction. Therefore, in the following the Coulomb gauge will be adopted. In such a framework, $\phi$ is related to the electrostatic component of the electric field , and to the longitudinal part of the electric field. On the other hand, $\mathbf{A}$ is related to the magnetic field, and to the transverse component of the electric field. We remind the reader that the electric and magnetic fields can be expressed in terms of scalar and vector potentials as $\mathbf{E} = -\mathbf{\nabla}\phi-\frac{1}{c}\partial_{t}\mathbf{A}$ and $\mathbf{B} = \nabla \times \mathbf{A}$. 
In Coulomb Gauge, the longitudinal part of the field only depend on $\phi$. Under this condition the Gauss law for the electric field ($\nabla\cdot\mathbf{E}= 4\pi\rho $ where $\rho$ is the electron density ) becomes the Poisson's equation ($ -\mathbf{\nabla}^{2}\phi = 4\pi \rho $) 
and therefore:
\begin{equation}
   \frac{1}{8\pi} \int d\mathbf{r}   \phi \nabla^2 \phi  = -\frac{1}{2} \int d\mathbf{r} \phi \rho (\mathbf{r} , t).
\end{equation}
Collecting this term with $\int \phi \rho d\mathbf{r}$ in Eq~\ref{eq:H}, gives the well known instantaneous electron-electron Coulomb repulsive contribution $\frac{1}{2} \I \int d\mathbf{r}' \frac{\rho_{e}(\mathbf{r}, t)\rho_{e}(\mathbf{r}', t)}{\lvert \mathbf{r} - \mathbf{r' \rvert}}$. \\

To facilitate the quantization (see next section), it is usually convenient expressing Hamiltonian~\ref{eq:H} in terms of so-called normal variables. These variable are defined in a similar way to the "ladder operator" method used to solve the quantum harmonic oscillator~\cite{ZwiebachQM2022, CohenTanoudjiQED}. To begin with, we rewrite Maxwell's equation in reciprocal space in the following way:
\begin{align}
    & \partial_{t}\mathcal{E}_{\mathrm{trans.}}(\mathbf{k}, t) = ic \mathbf{k} \times \mathcal{B} (\mathbf{k}, t) -4\pi \mathcal{J}_{\mathrm{trans.}}(\mathbf{k}, t) \\
    & \partial_{t}\mathcal{B}(\mathbf{k}, t) = -i\mathbf{k} \times \mathcal{E}_{\mathrm{trans.}}(\mathbf{k}, t). 
\end{align}
From the previous equation, one notices the following relationship when $ \mathcal{J}_{\mathrm{trans.}} = 0$:
\begin{equation}
    \partial_{t}\left( \mathcal{E}_{\mathrm{trans.}} \pm c \boldsymbol{\kappa}  \times \mathcal{B} \right) = \pm i \omega \left( \mathcal{E}_{\mathrm{trans.}} \pm c \boldsymbol{\kappa}  \times \mathcal{B}  \right)
\end{equation}
where $\omega = c |\mathbf{k}|$ and $\boldsymbol{\kappa} = \frac{\mathbf{k}}{|\mathbf{k}|}$. From the previous equation, it appears natural to introduce two new variables, even if $ \mathcal{J}_{\mathrm{trans.}} \neq 0$:
\begin{align}
    & \vec{\zeta} (\mathbf{k} , t) = - i \frac{\sqrt{2\pi \hbar \omega}}{2} \left[ \mathcal{E}_{\mathrm{trans.}} (\mathbf{k}, t) - \boldsymbol{\kappa} \times \mathcal{B} (\mathbf{k}, t) \right] \\
    & \vec{\zeta}^{*} (-\mathbf{k} , t) = - i \frac{\sqrt{2\pi \hbar \omega}}{2} \left[ \mathcal{E}_{\mathrm{trans.}} (\mathbf{k}, t) + \boldsymbol{\kappa} \times \mathcal{B} (\mathbf{k}, t) \right]
\end{align}
Using the previous relationship one is able to express the electric and magnetic field in terms of normal variables:
\begin{align} \label{eq:EB_normal}
    & \mathcal{E}_{\mathrm{trans.}} = i \sqrt{ 2\pi \hbar \omega } \left( \vec{\zeta} (\mathbf{k} , t)  - \vec{\zeta}^{*} (-\mathbf{k} , t) \right) \\
    & \mathcal{B}_{\mathrm{trans.}} = i \sqrt{ 2\pi \hbar \omega } \left( \boldsymbol{\kappa} \times \vec{\zeta} (\mathbf{k} , t)  + \boldsymbol{\kappa} \times \vec{\zeta}^{*} (-\mathbf{k} , t) \right) 
\end{align}
$\mathbf{E}_{\mathrm{trans.}}$ and $\mathbf{B}$ can be obtained using a Fourier transform.

Using these normal variables, the Parseval-Plancherel identity~\cite{CohenTanoudjiQED} allows writing the following:
\begin{equation}
    \frac{1}{8\pi} \int d^{3}r \left(  \mathbf{E}_{\mathrm{trans.}}^{2} + \mathbf{B}^{2} \right) = \frac{1}{8\pi}  \int d\mathbf{k} \left(  \lvert \mathcal{E}_{\mathrm{trans.}} \rvert^{2} + \lvert \mathcal{B} \rvert^{2} \right)
\end{equation}
and changing $\mathbf{k} \rightarrow - \mathbf{k}$ in the second term of the right-hand side of Eq.\ref{eq:EB_normal} we can rewrite the electromagnetic field Hamiltonian as: 
\begin{equation}\label{eq:FieldHam}
    \frac{1}{8\pi} \int d^{3}r  \left(  \mathbf{E}_{\mathrm{trans.}}^{2} + \mathbf{B}_{\mathrm{trans.}}^{2} \right) = \int d^{3}k \frac{\hbar \omega} {2} \left[ \vec{\zeta}^{*} \vec{\zeta} + \vec{\zeta} \vec{\zeta}^{*} \right]
\end{equation}
where the short notation $\vec{\zeta} = \vec{\zeta} (\mathbf{k}, t)$ has been used. \\
The vector potential $\mathbf{A}$ can be expressed in terms of the normal variables 
\begin{equation}\label{eq:vecpot_normalvar}
     \mathbf{A} (\mathbf{r}, t) = \int d^{3}k \sqrt{\frac{2 \pi \hbar c}{\omega\left( 2 \pi \right) ^{3}}} \Big(  \vec{\zeta} 
     \exp(i \mathbf{k}\cdot \mathbf{r})+ \vec{\zeta}^{*} 
     \exp( - i \mathbf{k}\cdot \mathbf{r})  \Big)
\end{equation}
Substituting Eq.s~\ref{eq:FieldHam} and \ref{eq:vecpot_normalvar} in Eq.~\ref{eq:H} we obtain the following expression for the classical Hamiltonian:

\begin{widetext}
 \begin{eqnarray}\label{eq:H_normalvar}
      {H} && = \int d^{3}k \frac{\hbar \omega} {2} \left[ \vec{\zeta}^{*} \vec{\zeta} + \vec{\zeta} \vec{\zeta}^{*} \right] + \I \Psi_{e}^{\dagger}\left[c\alpha^{i} (-i\hbar \nabla_i -\frac{e}{c}A^{i} -\frac{e}{c}A_{\text{ext}}^{i}) +\beta m_{e}c^2 \right] \Psi_{e} \nonumber\\
        && + \frac{1}{2} \I \int d^{3}r^{'} \frac{\rho_{e}(\mathbf{r}, t)\rho_{e}(\mathbf{r}^{'}, t)}{\lvert \mathbf{r} - \mathbf{r}^{'} \rvert} + \int \phi_{ext} \rho d\mathbf{r}. 
 \end{eqnarray}  
\end{widetext}
\subsubsection{\label{sec:level2.3}  Hamiltonian Quantization}
 From now on, the Sch\"odinger picture is adopted and therefore all operators will be considered time-independent. A quantized form of Hamiltonian~\ref{eq:H_normalvar} can be obtained by promoting the normal variables ($\vec{\zeta}^*$/$\vec{\zeta}$) to the corresponding $\mathbf{k}$-dependent field operators ($a'^{\dagger}_{\vec{\epsilon}}$/$a'_{\vec{\epsilon}}$), where $\vec{\epsilon}$ is the polarization vector, exploiting the decomposition of $\vec{\zeta}^*$/$\vec{\zeta}$ on the basis described by both polarization vectors $\vec{\zeta} =\sum_{\epsilon}\zeta_{\epsilon}\vec{\epsilon}$ and $\vec{\zeta}^{*} =\sum_{\epsilon}\zeta_{\epsilon}^{*}\vec{\epsilon}$.:  
\begin{align}
    & \zeta_{\vec{\epsilon}}(\mathbf{k}, t) \longrightarrow a'_{\vec{\epsilon}}(\mathbf{k}) \nonumber \\
    & \zeta^{*}_{\vec{\epsilon}}(\mathbf{k}, t) \longrightarrow a'^{\dagger}_{\vec{\epsilon}}(\mathbf{k}) 
\end{align}
satisfying the following commutation relations:
\begin{equation}
    \left[ a'_{\vec{\epsilon}}(\mathbf{k}) , a'^{\dagger}_{\vec{\epsilon} '}(\mathbf{k'})  \right] = \delta_{\vec{\epsilon} \vec{\epsilon} '} \delta ( \mathbf{k} - \mathbf{k '}).
\end{equation}
 In terms of the field operator the vector potential becomes:
\begin{align}\label{eq:vecpot_quantum}
     \mathbf{A} (\mathbf{r}) &= \sum_{\vec{\epsilon}} \int d^{3}k \sqrt{\frac{2 \pi \hbar c}{\omega\left( 2 \pi \right) ^{3}}} \Big[  a'_{\vec{\epsilon}}(\mathbf{k})\exp(i \mathbf{k}\cdot \mathbf{r})\vec{\epsilon}\nonumber \\ 
     &+ a'^{\dagger} _{\vec{\epsilon}}(\mathbf{k})\exp( - i \mathbf{k}\cdot \mathbf{r})\vec{\epsilon} \Big].
\end{align}
In the confined space of the optical cavity the wave vector $\mathbf{k}$ assumes discrete values consequently defining a discrete spectrum of field modes characterized by the direction of the $\mathbf{k}$ vector and by the polarization of the field oscillations ($\vec{\epsilon}$). Ideally, one should consider the full sum, and include an infinite amount of modes, however, it is in practice unfeasible. Therefore, quite often, only one or a few modes are included explicitly while the rest is simply not considered. This approximation is usually acceptable if the molecular states are well separated compared to the magnitude of the Rabi-Splitting (\textit{cf} Fig:~\ref{fig:Rabi_splitting_scheme}). In the relativistic context, the single/few-modes approximation can also be applied, but it requires a little more attention on the selection of the modes that need to be selected/discarded.  There, we will identify the mode enhanced by the device (e.g cavity, circuit, plasmon, etc.) as \textbf{k}$_{\text{dm}}$. Doing so, we can rewrite the vector potential in such a way:  
\begin{equation}\label{eq:sum_split}
     \mathbf{A} (\mathbf{r}) = \mathbf{A}_{\text{dm}} (\mathbf{r}) + \mathbf{A}_{\text{om}} (\mathbf{r})   \\
\end{equation}
where:
\begin{align}\label{eq:Acav}
     \mathbf{A}_{\text{dm}} (\mathbf{r}) &=  \sum_{\vec{\epsilon}}\mathrm{A}(\mathbf{k}_{\text{dm}})\Big[  a'_{\vec{\epsilon}}(\mathbf{k}_{\text{dm}})\exp(i \mathbf{k}_{\text{dm}}\cdot \mathbf{r})\vec{\epsilon} \nonumber \\
     & +a'^{\dagger}_{\vec{\epsilon}} (\mathbf{k}_{\text{dm}})\exp( - i \mathbf{k}_{\text{dm}}\cdot \mathbf{r})\vec{\epsilon}\Big]. 
\end{align}
and 
\begin{equation}\label{eq:Ares}
     \mathbf{A}_{\text{om}} (\mathbf{r})  = \mathbf{A} (\mathbf{r}) - \mathbf{A}_{\text{dm}} (\mathbf{r})    \\
\end{equation}

where $\mathrm{A}(\mathbf{k_{\text{dm}}}) =  \sqrt{\frac{2 \pi \hbar c}{\omega_{\text{dm}}\left( 2 \pi \right) ^{3}}}$. Splitting the sum into two terms allows for the identification of the terms corresponding to the mode enhanced by the device and $\mathbf{A}_{\text{om}}$ corresponding to the remaining part of the vector potential containing all the other modes. Such a splitting of  the vector potential $\mathbf{A}$ is motivated by the fact that the $\int d\mathbf{r} \mathbf{j} \cdot \mathbf{A}$ term, where $\mathbf{j} = ec\Psi_{e}^{\dagger} \mathbf{\alpha} \Psi_{e}$, includes several effects. In particular, it includes at the same time the electron-photon interaction as well as many other known energy terms (current-current term, frequency dependent Breit term, vacuum polarization, etc.~\cite{DyallBook2007, ReiherBook, Liu_RQC_2017}). Such a re-writing allows us to handle separately the photons coupled to the molecular system due to the device and what are known as relativistic corrections. Doing so, we can exploit tools from both \abinit polaritonic quantum chemistry and relativistic quantum chemistry to evaluate the various contributions. 
Since the molecular systems are usually significantly smaller than the wavelength of the cavity field, we are entitled to apply the dipole approximation imposing that $\exp(i \mathbf{k}_{\text{dm}}\cdot \mathbf{r}) \sim 1 $ and $\mathbf{A}_{\text{dm}} (\mathbf{r}) \sim \mathbf{A}_{\text{dm}} (\vec{0})$ in Eq.~\ref{eq:sum_split}. This significantly simplifies the photonic part of the Hamiltonian. \smallbreak 
For the fermionic part, the spinor fields are promoted to spinor fields operators satisfying the following equal-time anti-commutation relations:
\begin{equation}
    \{\Psi(\vec{r})_{\mu}, \Psi^{\dagger}(\vec{r}')_{\nu}\} = \delta_{\mu \nu} \delta(\vec{r} - \vec{r}').
\end{equation}
The QED Hamiltonian can then be written as: 
\begin{align}\label{eq:H_QED}
    \mathrm{H}_{QED} &= \int d^{3} \Psi^{\dagger} h^{(1)}_{D} \Psi \nonumber \\
    &+ \frac{1}{2} \I d^{3}r' \Psi^{\dagger}(\mathbf{r}) \Psi(\mathbf{r}) \frac{1}{\lvert \mathbf{r} - \mathbf{r}^{'} \rvert} \Psi^{\dagger}(\mathbf{r}') \Psi(\mathbf{r}')\nonumber \\ 
    &+ \sum_{\vec{\epsilon}} \hbar \omega_{\text{dm},\vec{\epsilon}} \left( a'^{\dagger}_{\text{dm},\vec{\epsilon}}a'_{\text{dm},\vec{\epsilon}}+ \frac{1}{2}\right).
\end{align}
Where
\begin{align}
    h^{(1)}_{D} = & c \alpha^{i} \left( p_{i} -\frac{e}{c}A_{\text{dm},i}(\vec{0}) -\frac{e}{c}A_{\text{om},i}(\mathbf{r}) -\frac{e}{c}A_{\text{ext},i}(\mathbf{r})\right) \nonumber \\
    & + \phi_{\text{ext}} + \beta m_{e} c^{2}.
\end{align}

Since we assume the single-mode approximation, the radiative term ($\sum_{\vec{\epsilon}} \hbar \omega_{\text{om},\vec{\epsilon}} \left( a'^{\dagger}_{\text{om},\vec{\epsilon}}a'_{\text{om},\vec{\epsilon}}+ \frac{1}{2}\right)$) involving the other modes only contributes in a trivial way. Therefore, in order to make the equations easier to read, we discarded this term which only represents a mere rigid shift of the energy levels of the system.

\subsubsection{Relativistic Pauli-Fierz Hamiltonian in the length gauge}\label{subsec:RPF}
When investigating molecular systems, it is usually more convenient to apply a unitary transformation to the field modes, allowing for a direct coupling between the field operators and the molecular dipole. This transformation is known as the length-gauge transformation:
\begin{equation}\label{eq:length_gauge_trans}
    U = \exp(\frac{ie}{\hbar c} \mathbf{A}_{\text{dm}}(\vec{0})\cdot \mathbf{R}) 
\end{equation}
where $\mathbf{R} = \I \Psi^{\dagger} \mathbf{r} \Psi$. Hamiltonian~\ref{eq:H_QED} can be transformed in the length-gauge form by application of Eq.~\ref{eq:length_gauge_trans}:
\begin{equation}\label{eq:transfo}
 H^{l}_{RPF} = U^{\dagger} H_{QED} U
\end{equation}
This step has to be followed by two other transformations. Firstly, a rotation which ensures that the photonic part is real, then a so-called coherent state transformation to remove the apparent origin dependence in the Hamiltonian~\cite{Flick_PRA_QED_DFT_2014, PhysRevB_2019_QED_Ronca, phys_rev_x_Ronca_2020} (more details on the derivation can be found in the section 1.3 in SI and in Ref.~\citenum{RokajJPhysB2018}). It is important to point out that the requirement for the photonic part to be real is not crucial in the relativistic case, but, for consistency with the non-relativistic works, we chose to keep it also in this context.  \\

Using these transformations we obtain:
\begin{align}\label{eq:H_QED_length_coherent}
    \mathrm{H}^{l}_{RPF} &= \I \Psi^{\dagger} h \Psi \nonumber \\
    &+ \frac{1}{2}\I d\mathbf{r}'\Psi^{\dagger}(\mathbf{r}) \Psi(\mathbf{r}) \frac{1}{\lvert \mathbf{r} - \mathbf{r}^{'} \rvert} \Psi^{\dagger}(\mathbf{r}') \Psi(\mathbf{r}')\nonumber \\
    & + \sum_{\vec{\epsilon}} \hbar \omega_{\text{dm}} \left( a^{\dagger}_{\vec{\epsilon}}a_{\vec{\epsilon}} +\frac{1}{2}\right) + \left(\frac{e}{c}\sqrt{\frac{2\pi}{V}}\vec{\epsilon}\cdot \left[ \mathbf{R} - \langle \mathbf{R} \rangle \right] \right)^{2} \nonumber \\
    &-\sqrt{\hbar \omega_{\text{dm}}}\left( \frac{e}{c} \sqrt{ \frac{2\pi}{V}}\vec{\epsilon}\cdot \left[ \mathbf{R} - \langle \mathbf{R} \rangle \right] \right) \left( a_{\vec{\epsilon}} + a^{\dagger}_{\vec{\epsilon}} \right),
\end{align}
Here, $a_{\text{dm},\vec{\epsilon}}$, $a_{\text{dm},\vec{\epsilon}}^{\dagger}$ have been relabeled to emphasize that, due to the length gauge transformation, the latter are modified. In fact, $\langle a^{\dagger}_{\text{dm},\vec{\epsilon}}a_{\text{dm},\vec{\epsilon}} \rangle$ does not coincide with the number of photons related to the mode \textbf{k}$_{\text{dm}}$ anymore~\cite{Flick_QED_DFT_ACS_2018}.
In Eq.~\ref{eq:H_QED_length_coherent}, the presence of the expectation value of the total dipole moments ($e\langle \mathbf{R} \rangle$) ensures that the Hamiltonian will not explicitly depend on the origin of the reference system for neutral molecules.
In atomic and molecular physics/chemistry, it is usually more convenient to expand the electronic field on a basis of atomic/molecular orbitals which are solutions of the Dirac equation in an external potential (the Coulomb potential of the nuclei, for instance). The field operator expanded on such a basis writes as:
\begin{equation}
    \hat{\Psi}_{e} (\vec{r}) = \sum_{p} \hat{c}_{p} \phi_{p} (\vec{r})  \text{ and } \hat{\Psi}^{\dagger}(\vec{r}') = \sum_{p} \hat{c}^{\dagger}_{p} \phi^{\dagger}_{p} (\vec{r}) 
\end{equation}
where the hat is used to indicate operators. $\phi_{p}$ are the solutions of the Dirac equation with an external potential. Notice that $\phi$ are mono-electronic/positronic spinors. In the following, the hat will be dropped to avoid an overload of symbols.  
The electronic creation ($c^{\dagger}$) and annihilation ($c$) operators satisfying anti-commutation rules:
\begin{equation}
    \{c_{p}, c^{\dagger}_{q} \} = \delta_{pq}.
\end{equation}
Using these operators the energy components of Hamiltonian~\ref{eq:H_QED_length_coherent} can be expressed in a second quantized form implementable in a quantum chemistry code. In the following, we will use a shorthand notation for the mono- and bi- electronic integrals that are defined as follows:
\begin{align}
    & O_{pq} = \I \phi^{\dagger}_{p} (\mathbf{r})O \phi_{q}(\mathbf{r})\\
    & (pq | rs) = \I d\mathbf{r}' \phi^{\dagger}_{p}(\mathbf{r})\phi^{\dagger}_{r} (\mathbf{r}') g(\mathbf{r}, \mathbf{r}')\phi_{s}(\mathbf{r}')\phi_{q}(\mathbf{r}) 
\end{align}
where $O$ is a generic one-body operator and $(pq|rs)$ are the well known two-electrons integrals.
Therefore, the second quantized (sq) Hamiltonian is:
\begin{equation}\label{eq:FinalHsq}
    \mathrm{H}_{RPF} = \mathrm{H}^{(1)} + \mathrm{H}^{(2)} + \mathrm{H}^{(3)} 
\end{equation}
where:
\begin{align}\label{eq:SplitFinalHamiltonian}
   &\mathrm{H}^{(1)} =  \sum_{pq} \underbrace{\left[h_{pq} + \frac{2\pi e^{2}}{V c^{2}} \left( Q_{pq} - 2\langle \epsilon_{\alpha} \cdot \mathbf{R} \rangle (\epsilon_{\alpha} \cdot \mathbf{r}_{pq}) \right)\right]}_{\tilde{h}_{pq}}c_{p}^{\dagger} c_{q} \nonumber \\ 
     &\mathrm{H}^{(2)} = \frac{1}{2} \sum_{pqrs} \underbrace{\left[ (pq|rs) + 2 \times \frac{2\pi e^{2}}{V c^{2}} (\epsilon_{\alpha} \cdot \mathbf{r}_{pq})(\epsilon_{\alpha} \cdot \mathbf{r}_{rs}) \right]}_{(pq\widetilde{|}rs)}c^{\dagger}_{p}c^{\dagger}_{r}c_{s}c_{q} \nonumber \\
    &\mathrm{H}^{(3)} =  \sum_{\vec{\epsilon}} \hbar \omega_{\text{dm}} ( N_{\vec{\epsilon}} + \frac{1}{2}) + \frac{2\pi e^{2}}{V c^{2}} \langle \vec{\epsilon} \cdot \mathbf{R} \rangle^{2}
\end{align}
where $Q$ refers to the molecular quadrupole and $N_{\vec{\epsilon}}=a^{\dagger}_{\vec{\epsilon}}a_{\vec{\epsilon}}$. \\

As previously mentioned, the interaction term $\I \mathbf{j}\cdot \mathbf{A}_{\text{om}}$ has to be evaluated in an approximated way. Currently, a lot of effort from the relativistic quantum chemistry community is devoted to develop accurate and efficient ways to evaluate the contributions emerging from such a term.

An extensive summary about how to include these effects can be found in Ref.s~\citenum{indelicato2017introduction, Indelicato_9_2017Liu, Liu_RQC_2017, ReiherBook, Matyus_2024_QED, Saue2022JCP}. In Bound State QED (BSQED), this term is evaluated using techniques from Quantum Field Theory (QFT), which, however, represents a tremendous effort. Such formalism, while providing remarkable accuracy for atomic systems, is not really applicable to molecular electronic structure calculations. Another strategy is the use of effective potentials (Breit potential~\cite{Grant2006,ReiherBook, dyall2007introduction}, Uehling~\cite{Uehling1935, Grant2006, ReiherBook, Liu_RQC_2017}, Wichmann-Kroll~\cite{Wichmann1956PR,Liu_RQC_2017} potential and Self-Energy contributions~\cite{Pyykko_2003, Flambaum_PRA_2005,ReiherBook, Liu_RQC_2017}). Such strategy has been rather successful in evaluating full-Breit contribution and QED corrections with a satisfying accuracy. This approach represents the most convenient strategy to be adopted in order to compare these effects with the polaritonic ones. At this point, we would like to emphasize that such effects are fairly small, and Gaunt/Breit and QED corrections are mostly important for core orbitals where polaritonic mostly influence the valence orbitals. Nonetheless, in some cases, their effect on molecular properties such as ionization potential or electron affinity can be of comparable magnitude ($\sim0.01-0.05$~eV)~\cite{PastekaPRLPhysRevLett118, DePrince_JCP_2021_QED_CC_IP_EA, JCP_Rosario_2022_IP}. This point might require some carefulness and be further studied in the future. 
Notice that Hamiltonian~\ref{eq:FinalHsq} allows both for positive and negative energy states. This point will be discussed in details in the next section.
\subsubsection{Treatment of the negative energy states}\label{sec:positrons}
Physically, the negative energy solutions correspond to positronic states~\cite{DyallBook2007}. This reinterpretation becomes clearer if we split the sum over all state into a sum over the positive energy states and another over the negative energy ones. In this picture, the negative energy electron creation (annihilation) operators are reinterpreted as positive energy positron annihilation (creation) operators (see Ref.s~\citenum{DyallBook2007, Dyall2012, Watanabe_2023, LiuComment2024}). In this framework, the field operator can be decomposed in this way:

\begin{align}
   & \Psi_{e} (\vec{r}) = \sum_{p} \{ c_{p} \phi_{p} (\vec{r}) +  b^{\dagger}_{p} \psi_{-p}(\vec{r})\}  \\
   & \Psi^{\dagger}_{e} (\vec{r}) = \sum_{p} \{ c^{\dagger}_{p} \phi^{\dagger}_{p} (\vec{r}) +  b_{p} \psi^{\dagger}_{-p}(\vec{r})\} \\
   &  \{ c_{p}, c^{\dagger}_{q} \} = \delta_{pq} \text{ and } \{b_{p}, b^{\dagger}_{q}  \} = \delta_{pq} 
\end{align}
where $b_{p}, b^{\dagger}_{p}$ are respectively positron annihilation and creation operators. Using these operators, Hamiltonian~\ref{eq:FinalHsq} can be rewritten in terms of both electronic and positronic contributions:
\begin{equation}\label{eq:FinalHsqEP}
    \mathrm{\bar{H}}_{RPF} = \mathrm{\bar{H}}^{(1)} + \mathrm{\bar{H}}^{(2)} + \mathrm{H}^{(3)} 
\end{equation}
where:
\begin{equation}
    \mathrm{\bar{H}}^{(1)} = \sum_{p, q} \left\{c_{p}^{\dagger} c_{q} \tilde{h}_{pq} + c_{p}^{\dagger} b^{\dagger}_{q} \tilde{h}_{p\qbar} + b_{p}c_{q}\tilde{h}_{\pbar q} - b^{\dagger}_{q}b_{p}\tilde{h}_{\pbar \qbar} \right\}
\end{equation}
\begin{align}
    \mathrm{\bar{H}}^{(2)} =  \frac{1}{2} \sum_{p,q,r,s} & \{ c^{\dagger}_{p}c^{\dagger}_{r}c_{s}c_{q} (pq\widetilde{|}rs) + 2 (pq\widetilde{|}r\sbar)c^{\dagger}_{p}c^{\dagger}_{r}b^{\dagger}_{s}c_{q} \nonumber \\
    &+ 2(pq\widetilde{|}\rbar s)c^{\dagger}_{p}b_{r}c_{s}c_{q} - 2 (pq\widetilde{|}\rbar \sbar) c^{\dagger}_{p} b^{\dagger}_{s}b_{r} c_{q}  \nonumber \\
    &+ 2 (p\qbar\widetilde{|}\rbar s)c^{\dagger}_{p}b^{\dagger}_{q}b_{r}c_{s} +(p \qbar \widetilde{|} r \sbar) c^{\dagger}_{p} c^{\dagger}_{r}b^{\dagger}_{s}b^{\dagger}_{q} \nonumber \\
    &+ (\pbar q \widetilde{|} \rbar s)b_{p}b_{r}c_{s}c_{q} -2(\pbar \qbar \widetilde{|} \rbar s)b^{\dagger}_{q}b_{p}b_{r}c_{s} \nonumber \\
    &-2(\pbar \qbar \widetilde{|}r \sbar) c^{\dagger}_{r}b^{\dagger}_{s}b^{\dagger}_{q}c_{p} + (\pbar \qbar \widetilde{|} \rbar \sbar) b^{\dagger}_{s}b^{\dagger}_{q}b_{p}b_{r} \}
\end{align}
$\mathrm{H}^{(3)}$ and the modified one and two-electrons integrals, in Eq.~\ref{eq:FinalHsqEP} have been defined in Eq.~\ref{eq:SplitFinalHamiltonian}. Barred indices correspond to positronic indices.
In this article we do not address the role of the fermionic vacuum that is at the heart of many discussions~\cite{Dyall2012, Watanabe_2023, LiuComment2024}. Indeed, the choice of the Hamiltonian and the possible reinterpretation of negative energy states has an influence on the nature of the fermionic vacuum which in turn has some consequence on the expectation value of the Hamiltonian. Fortunately, at the Hartree-Fock level there is no real dependence on the vacuum nature (however, it becomes important when developing correlated methods).

\subsection{The Hartree-Fock approximation}
Since it treats explicitly all the interactions between relativistic electrons/positrons and the photons of the cavity field, Hamiltonian~\ref{eq:FinalHsqEP} represents the perfect starting point for the development of \abinit theories suitable to simulate molecular systems containing heavy atoms in optical devices. 
In this section, we will use Hamiltonian~\ref{eq:FinalHsqEP} to develop the first relativistic 4-components Hartree-Fock (HF) approach for polaritonic chemistry. In quantum chemistry, HF represents the simplest approximation respecting the right symmetry of all the involved particles. HF also provides access to physically meaningful sets of atomic/molecular orbitals that can be subsequently used to develop more accurate correlated theories. In the present paper $\phi_{\text{ext}}$ has been replaced by the (point charge) Coulomb potential of the nuclei in the Born-Oppenheimer approximation and \textbf{A}$_{\text{ext}}=0$ in Hamiltonian~\ref{eq:FinalHsqEP}. In future work the point charge nuclear potential will be replaced by a finite size model, but from a formal point of view, it does not introduce any major change to the equations. The two electrons integrals correspond to the instantaneous Coulomb interaction. 
HF is a mean field theory, hence the involved quantum species do not directly interact between each other but travels in space under the action of the average potential of the other particles. For the purely electronic case, this implies that the wave function is described by a single Slater determinant of atomic/molecular orbitals or spinors in the relativistic case:
\begin{equation}\label{eq:determinantal_wf}
\Phi (\mathbf{r}_1,...,\mathbf{r}_N) = \frac{1}{\sqrt{N!}}
\begin{vmatrix}
\phi_1(\mathbf{r}_1) & \phi_1(\mathbf{r}_2) & \cdots & \phi_1(\mathbf{r}_N) \\
\phi_2(\mathbf{r}_1) & \phi_2(\mathbf{r}_2) & \cdots & \phi_2(\mathbf{r}_N) \\
\vdots & \vdots & \ddots & \vdots \\
\phi_N(\mathbf{r}_1) & \phi_N(\mathbf{r}_2) & \cdots & \phi_N(\mathbf{r}_N)
\end{vmatrix}.
\end{equation}
This ansatz ensures the anti-symmetry of the wave function under particle exchange.
In this case, three kinds of particles are involved: electrons, positrons and photons, each one respecting its own statistics (fermionic for electrons/positrons and bosonic for photons).
The wave-function can be written as the product of distinct wave functions for every species:
\begin{equation}\label{eq:DHF_ansatz}
    \Psi = \Phi_{e} \otimes \Phi_{p} \bigotimes_{\tau}|0\rangle_{\tau}
\end{equation}
where $\Phi_{e}$ and $\Phi_{p}$ are single determinants for electronic and positronic spinors respectively,
$|0\rangle_{\tau}$ is the vacuum state associated to the photonic mode $\tau = (k, \vec{\epsilon} )$ and $\bigotimes_{\tau}\ket{0}_{\tau}$ is a shorthand notation for $\prod_{\tau} \otimes \ket{0}_{\tau}$. A similar ansatz has been already applied in the non-relativistic version of the QED-HF approach, and its implications are discussed in detail in Ref.~\citenum{phys_rev_x_Ronca_2020}. 
Ansatz~\ref{eq:DHF_ansatz} can be rewritten, for practical reasons, in terms of second quantized electronic and positronic operators as:
\begin{equation}
    \Psi = (\prod_{i=0}^{n} c^{\dagger}_{i} |0\rangle_e) \otimes (\prod_{j=0}^{m}b^{\dagger}_{j}|0\rangle_p)\bigotimes_{\tau}|0\rangle_{\tau}
\end{equation}
where $|0\rangle_e$ and $|0\rangle_p$ are the electronic and positronic vacuum states.

Projecting Hamiltonian~\ref{eq:FinalHsqEP} on Eq.~\ref{eq:DHF_ansatz} we can calculate the relativistic polaritonic HF energy:

\begin{align}\label{eq:HF_expval}
    \langle E \rangle_{\text{DHF}} &= \sum_{p}\tilde{h}_{pp} - \sum_{\pbar}\tilde{h}_{\pbar \pbar} + \frac{1}{2}\sum_{p,q} \{ (pp\widetilde{|}qq) - (pq\widetilde{|}qp) \} \nonumber \\
    &+ \frac{1}{2}\sum_{\pbar,\qbar} \{ (\pbar \pbar\widetilde{|}\qbar \qbar) - (\pbar \qbar\widetilde{|}\qbar \pbar) \} - \sum_{p, \qbar} \{ (pp\widetilde{|}\qbar \qbar) - (p\qbar \widetilde{|} \qbar p) \} \nonumber \\
    &+ \sum_{\alpha}\hbar \omega_{\alpha} (N_{\alpha} + \frac{1}{2}) + \langle h_{\text{nuc}} \rangle + \frac{2\pi e^{2}}{V c^{2}}\langle \epsilon_{\alpha} \cdot \mathbf{R} \rangle^{2}
\end{align}
where :
\begin{align}
    & \tilde{h}_{pp} = h_{pp} - \frac{2\pi e^{2}}{V c^{2}}\left[Q_{pp} - 2 \langle \epsilon_{\alpha} \cdot \mathbf{R} \rangle (\epsilon_{\alpha} \cdot \mathbf{r}_{pp})  \right] \\
    & \tilde{h}_{\pbar \pbar} = h_{\pbar \pbar} - \frac{2\pi e^{2}}{V c^{2}}\left[Q_{\pbar \pbar} - 2 \langle \epsilon_{\alpha} \cdot \mathbf{R} \rangle (\epsilon_{\alpha} \cdot \mathbf{r}_{\pbar \pbar})  \right] \\
    & (pp\widetilde{|}qq) = (pp|qq) - 2 \times \frac{2\pi e^{2}}{V c^{2}} \left((\epsilon_{\alpha} \cdot \mathbf{r}_{pp})(\epsilon_{\alpha} \cdot \mathbf{r}_{qq}) \right) \\
     & (pq\widetilde{|}qp) = (pq|qp) - 2 \times \frac{2\pi e^{2}}{V c^{2}} \left((\epsilon_{\alpha} \cdot \mathbf{r}_{pq})(\epsilon_{\alpha} \cdot \mathbf{r}_{qp}) \right) \\
       & (\pbar \pbar\widetilde{|}\qbar \qbar) = (\pbar \pbar|\qbar \qbar) - 2 \times \frac{2\pi e^{2}}{V c^{2}} \left((\epsilon_{\alpha} \cdot \mathbf{r}_{\pbar \pbar})(\epsilon_{\alpha} \cdot \mathbf{r}_{\qbar \qbar}) \right) \\
     & (\pbar \qbar\widetilde{|}\qbar \pbar) = (\pbar \qbar|\qbar \pbar) - 2 \times \frac{2\pi e^{2}}{V c^{2}} \left((\epsilon_{\alpha} \cdot \mathbf{r}_{\pbar \qbar})(\epsilon_{\alpha} \cdot \mathbf{r}_{\qbar \pbar}) \right) \\
      & (pp\widetilde{|}\qbar \qbar) = (pp|\qbar \qbar) - 2 \times \frac{2\pi e^{2}}{V c^{2}} \left((\epsilon_{\alpha} \cdot \mathbf{r}_{pp})(\epsilon_{\alpha} \cdot \mathbf{r}_{\qbar \qbar}) \right) \\
     & (p\qbar\widetilde{|}\qbar p) = (p\qbar|\qbar p) - 2 \times \frac{2\pi e^{2}}{V c^{2}} \left((\epsilon_{\alpha} \cdot \mathbf{r}_{p\qbar})(\epsilon_{\alpha} \cdot \mathbf{r}_{\qbar p}) \right) 
\end{align}
Notice that at the HF level, all terms that do not conserve the number of particles will have a zero expectation value. This is the case for the so-called bilinear term in Eq.s~\ref{eq:H_QED_length_coherent}
 which involves only one annihilation (creation) operator and therefore changes the number of photon. Consequently, this term gives a zero contribution to the ground state energy.

\subsubsection{The Fock-operator}
As for the bare electrons case, the HF problem can be solved by minimizing Eq.~\ref{eq:HF_expval} for variations in the spinors. This can be done by performing a rotation of the mono-electronic/positronic orbitals using the following operator:
\begin{equation}
  \mathrm{K} = \sum_{pq} \mathrm{K}_{pq} c^{\dagger}_{p}c_{q}
\end{equation}
and then imposing the stationary condition:
\begin{equation}
    \frac{\partial \langle E \rangle_{\text{DHF}}}{\partial \mathrm{K}_{pq}}=0.
\end{equation}
From this minimization we can obtain, following standard orbital rotation techniques~\cite{MolecularElecStruc2000, dyall2007introduction, Dyall2012}, a new Fock operator with matrix elements:
\begin{equation}\label{eq:Fock}
    f_{pq} = \tilde{h}_{pq} + \sum_{k}\left[ (pq\widetilde{|}kk) - (pk\widetilde{|}kq) \right] - \sum_{\kbar} \left[(pq\widetilde{|}\kbar \kbar) - (p\kbar\widetilde{|}\kbar q) \right]
\end{equation}
where $p,q$ are either positron or electron indices. 

The Fock operator in Eq.~\ref{eq:Fock} can be applied in a Roothan-Hall like procedure~\cite{MolecularElecStruc2000}:
\begin{equation}
    \mathbf{F}\mathbf{C}=\boldsymbol{\varepsilon}\mathbf{C}
\end{equation}
to optimize the orbital coefficients.
It is important to highlight that exactly as in the non-relativistic QED-HF  approach (see Ref.~\citenum{phys_rev_x_Ronca_2020} the Fock matrix in Eq.~\ref{eq:Fock} suffers from an explicit origin dependence if charged systems need to be investigated. This problem can be solved applying the so-called Strong Coupling (SC) HF approximation proposed in Ref.~\citenum{nature_riso_2022}. In this paper, we will focus, for the moment, on neutral molecular system and an SC extension of our method will be the topic of a future work. 

Due to the presence of negative energy states in the Hamiltonian spectrum, the spinor optimization procedure is in fact a minimax problem~\cite{dyall2007introduction, ReiherBook}. In our approach, equivalent results can be obtained when no explicit positron is included in the wave function ansatz:
\begin{equation}\label{eq:DHF_ansatz_short}
    \Psi = \Phi_{e} \bigotimes_{\tau}|0\rangle_{\tau}.
\end{equation}
consequently, the projection of Hamiltonian~\ref{eq:FinalHsqEP} on such wave functions will cancel all positronic dependent terms in Eq.~\ref{eq:HF_expval}.

This ansatz is the natural extension of the one usually used in non-relativistic QED-HF~\cite{phys_rev_x_Ronca_2020,nature_riso_2022}.
In this paper, this simplified ansatz will be used to generate the results presented in the results section. The positronic degrees of freedom will be included instead explicitly in a future implementation of the method.
\subsubsection{Kinetic Balance in presence of the field}\label{sec:kinetic_balance}
In relativistic quantum chemistry, the spinors can be expressed as two-components objects:
\begin{equation}
    \begin{bmatrix}
    \Psi_{1} \\
    \Psi_{2} \\
    \Psi_{3} \\
    \Psi_{4} \\
    
    \end{bmatrix} = \begin{bmatrix}
        \Psi^{L} \\
        \Psi^{S}
    \end{bmatrix}
\end{equation}
where $\Psi^{L}$ and $\Psi^{S}$ are called respectively "Large" and "Small" components.
In practice, the spinor solutions are expanded on a basis-set: $\Psi^{T} = \sum_{\mu}C^{T}_{\mu}\chi^{T}_{\mu}$ where $T = \{L, S\}$ and $\chi^{T}$ is a 2-spinor. In order to avoid the so-called variational collapse of the solution, a constraint on the large and small components of the basis set is applied~\cite{Grant2006,ReiherBook}. In the presence of a vector potential, such constraint has the following form:
\begin{equation}
    \chi^{S} = \frac{\mathbf{\boldsymbol{\sigma} \cdot \boldsymbol{\pi}}}{2 m_{e}c}\chi^{L}
\end{equation}
where $\boldsymbol{\sigma}=(\sigma_{x}, \sigma_{y}, \sigma_{z})$ is the vector of the three Pauli matrices and $\boldsymbol{\pi}=\mathbf{p} -\frac{e}{c}\mathbf{A}$ is the generalized momentum. Such prescription is called the magnetic balance condition~\cite{Komorsky2008_magneticbalance, Aucar_Saue_1999_Diamgn, PhysRevAKutz2003, LiuJCP2007}. In standard relativistic quantum chemistry, the momentum is simply $\mathbf{p}$.  We refer to this condition as the kinetic balance~\cite{Grant2006, ReiherBook}. 
In our case, since the generalized momentum is \textit{a priori} depending on the field, the magnetic balance needs, in principle, to be satisfied. This could require to have a field-dependent small-component basis-set. 
However, in the present context, we have shown that in the dipole-approximation, applying the length gauge transformation, the momentum operator of the theory is thus transformed in the following way (see section 1.3 of the SI for a more detailed derivation): 
\begin{equation}
    \mathbf{p} -\frac{e}{c}\mathbf{A}(\vec{0}) \rightarrow \mathbf{p}.
\end{equation}
Consequently, the kinetic balance condition:
\begin{equation}
    \chi^{S} = \frac{\boldsymbol{\sigma} \cdot \mathbf{p}}{2 m_{e}c}\chi^{L}
\end{equation}
can be applied in this case. It is important to stress that this is only possible under the dipole-approximation, otherwise, the length gauge transformation would not provide such a simple expression for the Hamiltonian and the associated momentum.
\subsection{Excited states properties}
\begin{figure}[ht!]
    \centering
    \includegraphics[width=0.8\linewidth]{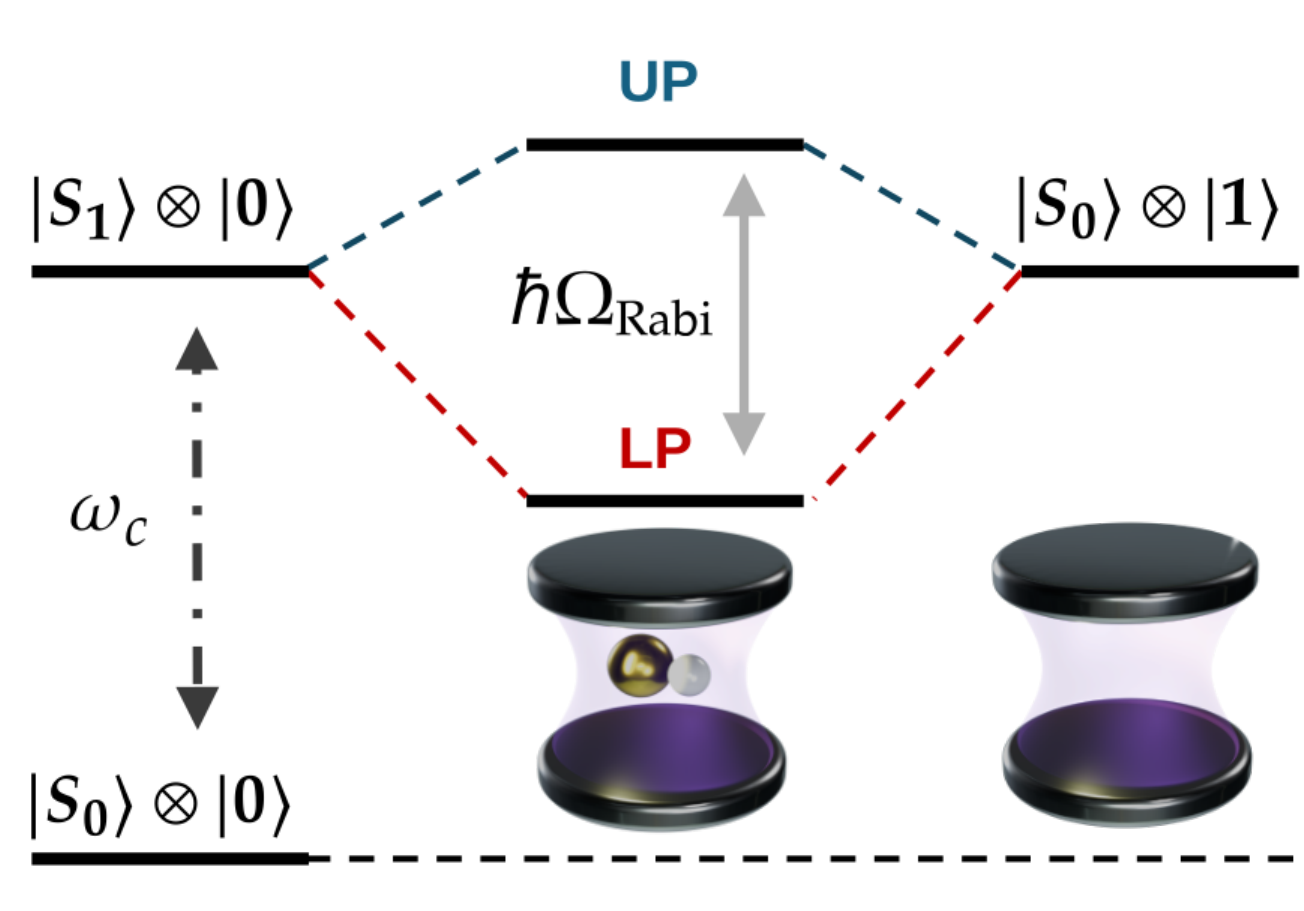}
    \caption{Scheme of a generic Rabi Splitting ($\hbar \Omega_{\text{Rabi}}$). UP and LP indicate the Upper and Lower polaritons respectively.}
    \label{fig:Rabi_splitting_scheme}
\end{figure}

In polaritonic chemistry, the signature property emerging from the strong coupling condition is the Rabi-splitting. It represents the energy separation between the polaritons formed by the mixing between matter and field states (see Fig.~\ref{fig:Rabi_splitting_scheme}).
Calculating Rabi-splittings, which could directly be compared with experimental data, requires access to the excited states of the coupled light matter system. At the HF level they can be simulated recurring to linear response theory.  
Recently, Castagnola et al.~\cite{Castagnola2024_linear_response} presented an  HF linear response theory for non-relativistic polaritonic systems. In this section, we present an extension of this approach to 4-components Dirac HF. Using linear response theory, excitations energies ($\omega_I$) can be obtained by solving the well known Casida equation~\cite{TDDFT_Casida1995}:
\begin{equation}
\label{eq:eigenCasida}
    \begin{bmatrix}
    A & B \\
    B^{*} & A^{*}
    \end{bmatrix}
    \begin{pmatrix}
    \vec{X_{I}} \\
    \vec{Y_{I}}
    \end{pmatrix} =
    \omega_{I} \begin{bmatrix}
    +1 & 0 \\
    0 & -1
    \end{bmatrix} 
    \begin{pmatrix}
    \vec{X_{I}} \\
    \vec{Y_{I}}
    \end{pmatrix}.
\end{equation}
The polaritonic expressions of the $A$ and $B$ matrices as derived in Ref.~\citenum{Castagnola2024_linear_response} assumes the form:
\begin{equation}
    A = 
    \begin{bmatrix}
    \omega_{\alpha}\delta_{\alpha \beta} & \sqrt{\omega_{\alpha}}(\mathbf{\lambda}_{\alpha}\cdot\mathbf{d}_{ib}) \\
    \sqrt{\omega_{\alpha}}(\mathbf{\lambda}_{\alpha}\cdot\mathbf{d^{*}}_{bi}) & \mathbf{A}_{el} 
    \end{bmatrix}
\end{equation}
\begin{equation}
    B = 
    \begin{bmatrix}
    0 & -\sqrt{\omega_{\alpha}}(\mathbf{\lambda}_{\alpha}\cdot\mathbf{d}_{ib}) \\
    -\sqrt{\omega_{\alpha}}(\mathbf{\lambda}_{\alpha}\cdot\mathbf{d^{*}}_{bi}) & \mathbf{B}_{el} 
    \end{bmatrix}
\end{equation}
where $\mathbf{A}_{el}$ and $\mathbf{B}_{el}$ have matrix elements:
\begin{equation}
 (\mathbf{A_{el}})_{ia,bj} = \delta_{ij}f_{ab} - \delta_{ab}f_{ij} + 2(a i\widetilde{|} b j) - (a b\widetilde{|}j i) 
\end{equation}
\begin{equation}
 (\mathbf{B_{el}})_{ia,bj} = (b i\widetilde{|}a j)  - 2(a i\widetilde{|} b j)  
\end{equation}
A similar approach, has been recently proposed by Konecny et al.~\cite{RelQEDDFT} to calculate excitation energies at the QED Dirac Kohn Sham level of theory based on previous work presented in the following references~\cite{Flick_QEDDFT_linear_response_2019,Yang_JCP_2021_QEDDFT_linear}. 
\section{\label{sec:level3}Results}
\subsection{Methods}
All the results presented in this section have been obtained from a development version of the PySCF software package~\cite{pyscf2018, pyscf2020}. All energies have been converged using default parameters in PySCF. In particular, the Pol-DHF approach has been implemented by applying modifications to the integrals used in standard DHF~\cite{QimingDHFJCTC2021}. By default the spinor integrals are evaluated on a basis set including both $j = l \pm \frac{1}{2}$. From a technical point of view, the Pol-DHF method scales as the standard DHF method (they only differ by some pre-factor).
So far, to simplify the interpretation of the results, we neglected the positrons in the treatment. A detailed discussion of the effects induced by the explicit inclusion of the positronic degrees of freedom will the topic of a future follow-up paper.
All calculations have been performed using the all-electron x2c-SVP basis-set~\cite{basissetexchange1996, basisetexchange2007, basissetexchange2019, x2cbasis} for the large component. The small component has been obtained via the kinetic balance prescription. The coupling value has been set to 0.05 a.u.. Though this coupling is not accessible with standard Fabry-Pérot cavity, it can be reached including collective effects or with other types of devices as in Ref~\citenum{USC_regime_RevModPhys_2019}. In this study, one molecule coupled to the device is considered with a high coupling value. However, it should not change the qualitive character of the effects observed~\cite{PhysRevX_Rosario_2023}. Morevover, including more molecules in the device can only be meaningful if electron-photon correlation is included in the theory~\cite{DePrince_JCP_2021_QED_CC_IP_EA, JCP_Rosario_2022_IP}. This will be the topic of a follow-up work. The datasets generated and analyzed during the current study can be reproduced using the PySCF development code, the inputs files and the geometries provided at the Zenodo link~\cite{zenodo}. 

\subsection{Ground state properties}\label{sec:GS_prop}
In this section, we analyze the field induced effects on the ground state properties of three metal hydrides (CuH, AgH and AuH). These complexes contain metals belonging to different periods of the 11$^\mathrm{th}$ group of the periodic table. Going down the groups, the velocity of the electrons (in particular of the inner ones) increases, approaching finite fractions of the speed of light in the gold case. In fact, gold complexes are well known to have interesting electronic properties due to the significant relativistic effects~\cite{Pykko2004, Gorin2007, pykko2008goldIII, BelpassiTheGOAT2008}.
In this paper, we will analyze how the polaritonic effects compete with the relativistic ones, generating modifications of the molecular electronic structure.
In Table~\ref{tab:DHF_GS_energies} we compare the total energies calculated at the DHF level of theory with those evaluated including the field (Pol-DHF).
\begin{table}[ht!]
    \centering
    \begin{tabular}{|c|c|c|c|}
        \hline
          molecule & DHF (Hartree) & Pol-DHF (Hartree) & $\Delta$E$_{\text{Pol}}$ (eV) \\
         \hline
          CuH &  -1653.11275 & -1653.11843   &  -0.1543 \\
         \hline
          AgH & -5338.68917 & -5338.69790 & -0.2374  \\
         \hline
          AuH & -21639.06979 & -21639.07709  & -0.1988 \\
         \hline
    \end{tabular}
    \caption{Ground state total energies calculated at the DHF and Pol-DHF level. $\Delta$E$_{\text{Pol}} = \text{E}_{\text{Pol-DHF}} - \text{E}_{\text{DHF}} $.  
    }
    \label{tab:DHF_GS_energies}
\end{table}
The field induced effects, despite small compared to the total energies, still represent a sizable (some tens of eV) variation on the energy of the system. In particular, it is interesting noticing that the field induced energy variation is quite similar for copper and gold (Cu - 0.15 eV, Au - 0.20 eV) while it is slightly bigger (in absolute value) for silver (Ag - 0.24 eV). This trend can actually be explained by the fact that under the QED-HF approximation, the only field term contributing to the total energy of the system is the dipole self-energy contribution $\left(\frac{e}{c}\sqrt{\frac{2\pi}{V}}\epsilon_{\alpha}\cdot \left[ \mathbf{R} - \langle \mathbf{R} \rangle \right] \right)^{2}$. As highlighted in Eq.~\ref{eq:SplitFinalHamiltonian} the one-electron term coming from the dipole self-energy directly depends on the molecular quadrupole, and it is well known from the literature (see Ref.~\citenum{KelloJCP1991}) that this as many other electronic properties (atomic radius, etc.)~\cite{SuzumuraIntJQuantumChem1999} show a very similar trend if we move down the group.

The effects observed on the total energy values are an indirect observation of the variations induced by the field on the system's molecular orbitals (MO). 
\begin{figure*}[ht!]
    \centering
    \includegraphics[width=0.32\linewidth]{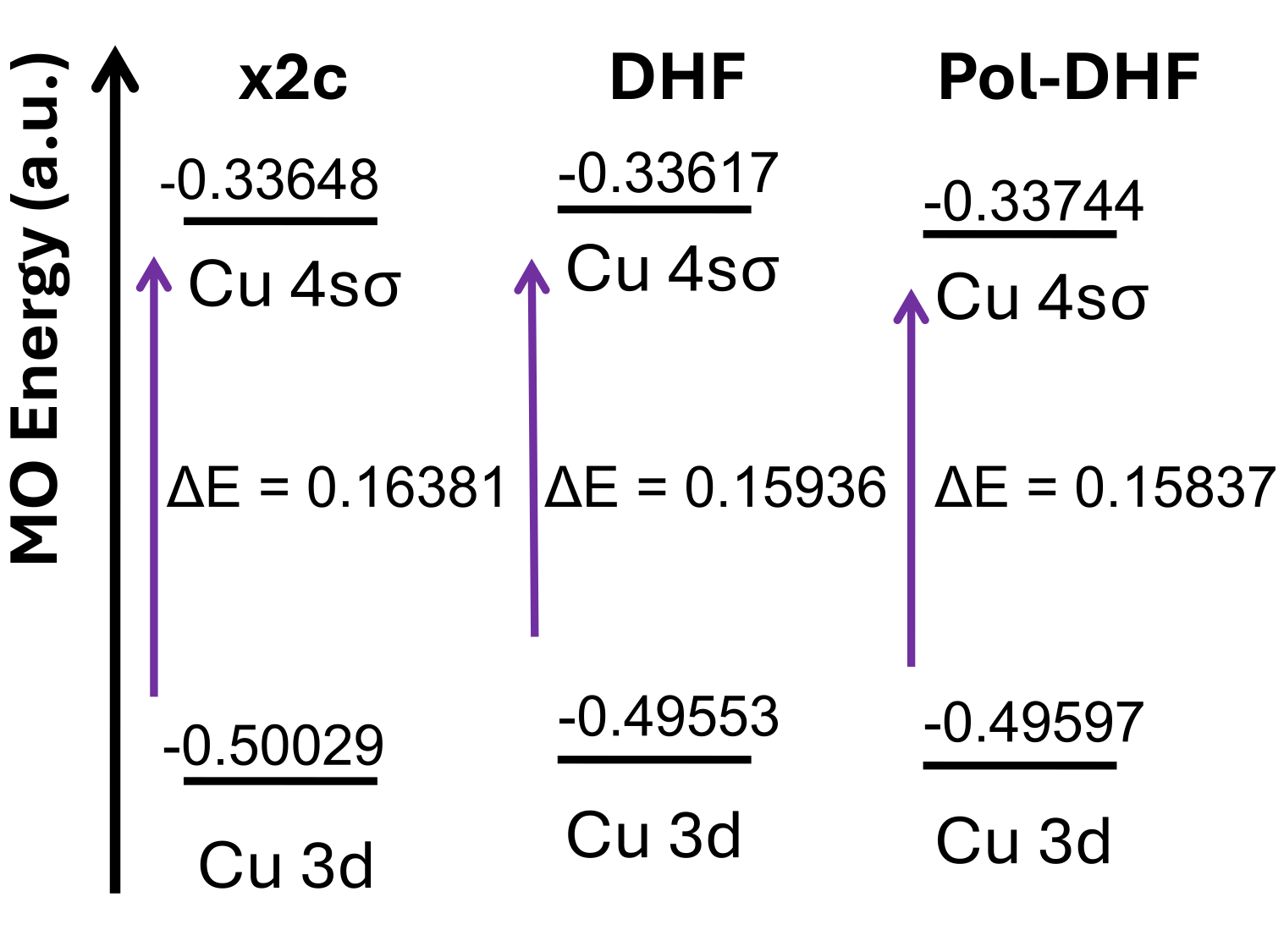}\hfill
    \includegraphics[width=0.32\linewidth]{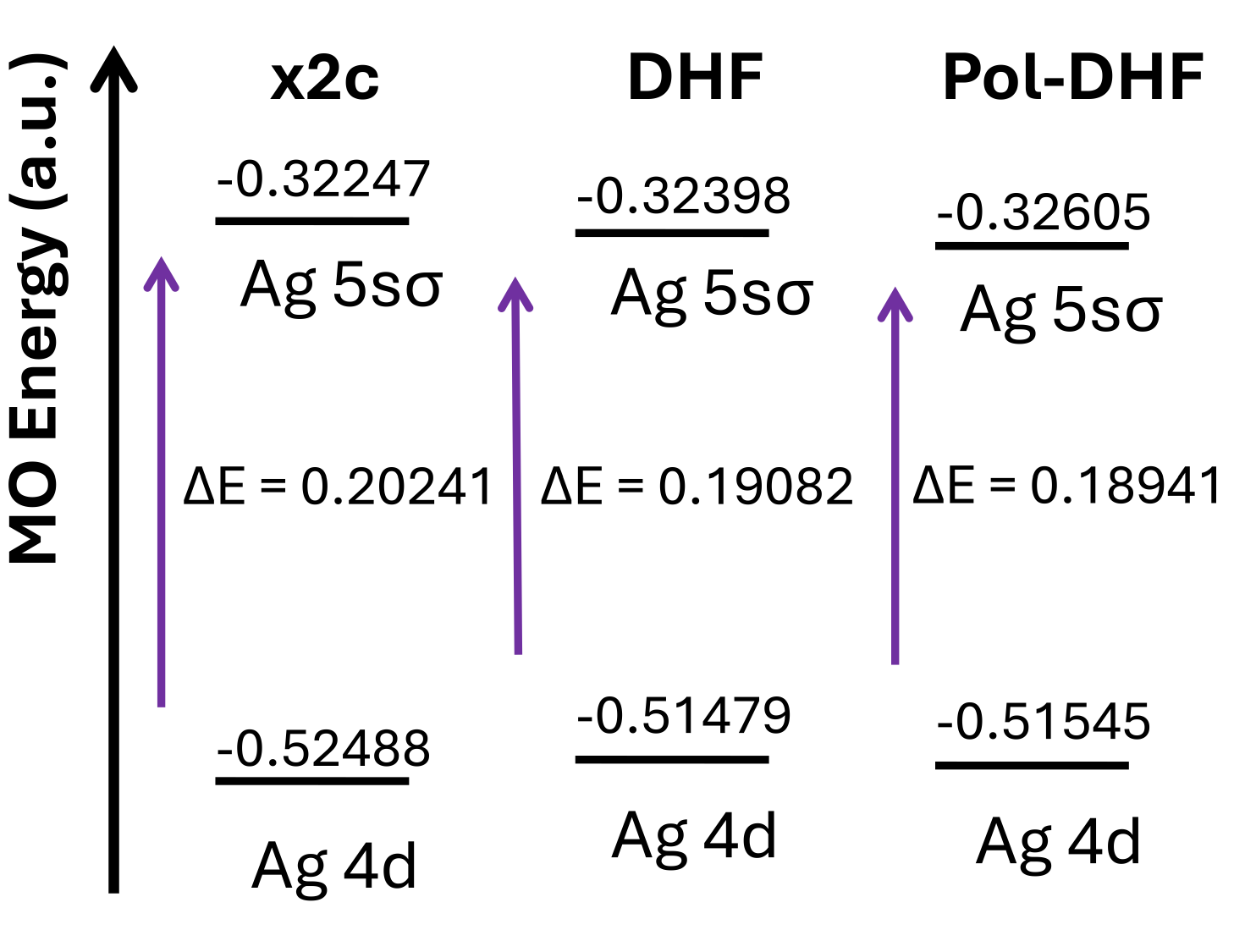}\hfill
    \includegraphics[width=0.32\linewidth]{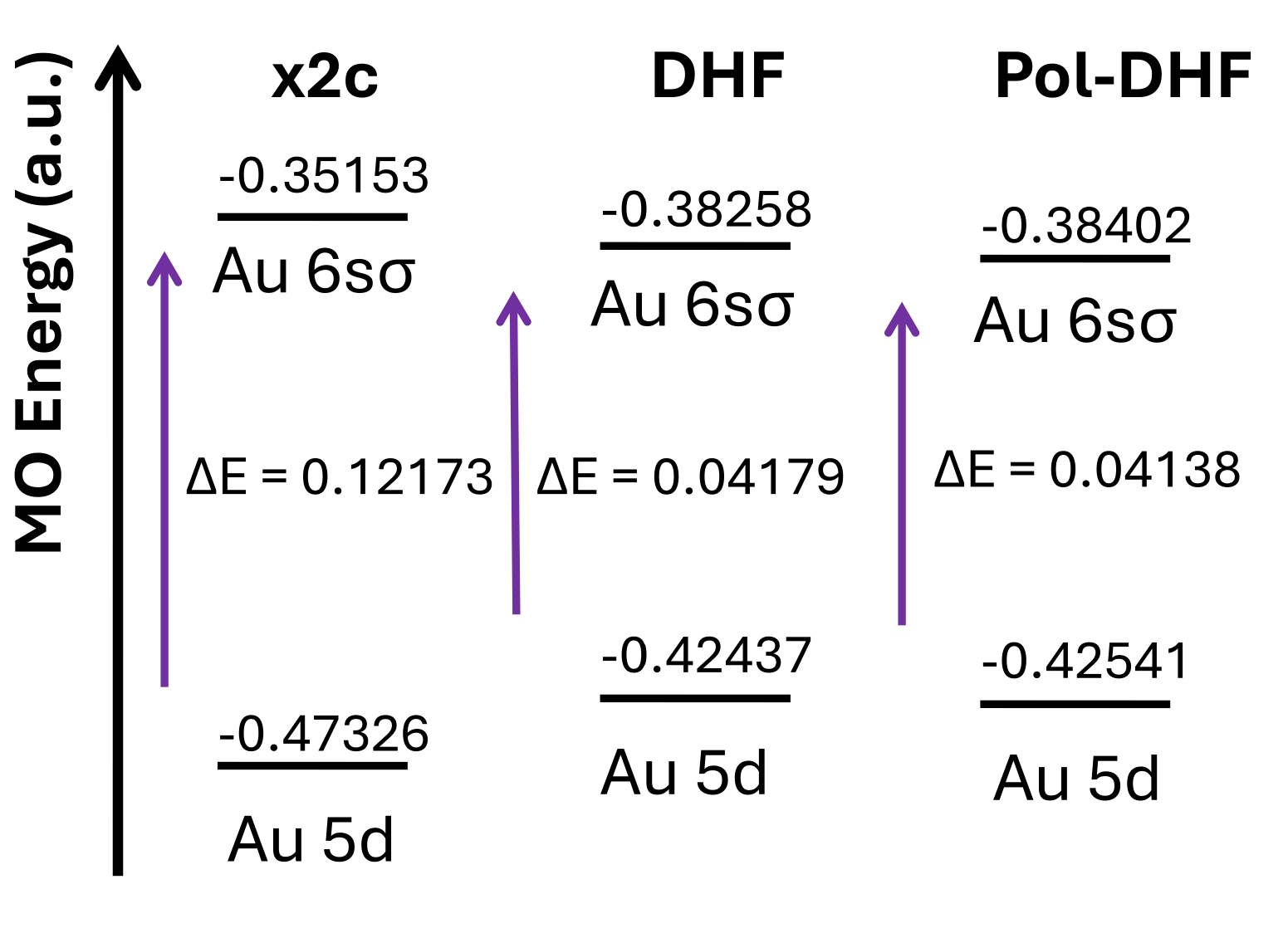}\hfill
    \caption{Change in energies from $nd \rightarrow (n+1)s\sigma$ in a.u.}
    \label{fig:levelsndns}
\end{figure*}
In Figure~\ref{fig:levelsndns}, we compare the orbital energies for the valence orbitals ($nd$ and $(n+1)s\sigma$) of the three complexes evaluated at the Pol-DHF level with those calculated at the DHF and spin-free eXact-2-component (SFX2C) level of theory~\cite{dyallx2c2001, pyscf2018, pyscf2020}.
As expected, the biggest variations in the orbital energies can be observed comparing  the spin free X2C results (as implemented in PySCF) with the DHF ones. In this case, the usual destabilization of the $d$ orbitals and consequent stabilization of the $s\sigma$ MOs can be observed. The reduction of the gap between these occupied MOs
decreases with the increasing relativistic character of the metal (smaller for Au than for Cu). The effect of the field on the electronic structure, though sizable, is much smaller.
\begin{table}[ht!]
    \centering
    \begin{tabular}{|c|c|c|c|}
        \hline
          molecule & $\Delta( \Delta E)$ (meV) & $\chi_{\text{E}}$ (\%) \\
         \hline
          CuH & 26.9  & 0.62 \\
         \hline
          AgH & 38.4 & 0.74 \\
         \hline
          AuH & 11.2  &  0.98\\
         \hline
    \end{tabular}
    \caption{The first column is the difference of the $nd \rightarrow (n+1)s\sigma$ gap in meV. The second column represents the relative change in energy due to Polaritonic effects.  
    }
    \label{tab:relenergies}
\end{table}
The field induced effects ($\Delta(\Delta E)$) on the $nd \rightarrow (n+1)s\sigma$ energy gap, barely appreciable in Figure~\ref{fig:levelsndns}, are reported in Table~\ref{tab:relenergies}. As already observed for the total energy, the largest field induced variation of the energy gap is observed for AgH. It is crucial, to point out that even in the gold case, where the effect is smaller, the cavity field is able to induce a $\sim$10 meV variation on the orbital gap. These values, represent a reasonable fraction ($\sim 0.25$ or higher) of a kcal/mol, relevant to observe variations of the chemical properties. 
It is important to highlight that the effects on the ground state properties we just presented are only accessible if the ultra-strong coupling regime~\cite{USC_regime_RevModPhys_2019}, like the one we simulated in our test analysis, can be reached. However, it is also important to remind that the cavity induced ground state effects could have a significant impact on the electronic and nuclear spin excitations (usually strongly affected by relativistic effects) also in more moderated coupling regimes. This aspect will be the topic of a following study.
 Interestingly the relative change induced by the field ($\chi_{\text{E}}=\frac{\Delta( \Delta E)}{\Delta E}*100$) is larger for AuH (about 1\% of the total gap) compared to the other systems (0.7\% for AgH and 0.6\% for CuH). This is a small difference and it is clearly not sufficient to draw conclusions. An extended study to verify if this trend still holds going down the periodic table could yield some interesting insights. Notice that these field induced variations of the $nd \rightarrow (n+1)s\sigma$ energy gap lead (for all systems) to a reduction of the gap due to a very small stabilization of the $(n+1)s\sigma$ MOs accompanied by a larger destabilization of the $nd$ orbitals.  
\subsubsection{Gaunt, Breit and polaritonic contributions to ground state energies}
In this work, we were able to include in a variational way the frequency independent Breit term:
\begin{equation}
  B(i, j) = - e^{2}\left\{ \underbrace{\frac{\mathbf{\alpha}_{i}\mathbf{\alpha}_{j}}{\lvert \mathbf{r}_{i} - \mathbf{r}_{j}\rvert}}_{\text{Gaunt}} + \underbrace{\frac{1}{2}(\mathbf{\alpha}_{i} \cdot \nabla_{i})(\mathbf{\alpha}_{j} \cdot \nabla_{j})\lvert\mathbf{r}_{i} - \mathbf{r}_{j}\rvert}_{\text{Breit}} \right\}
\end{equation}
and  performed a comparison between the energy variations introduced by the quantum field, with the Gaunt and Breit terms. 
\begin{table}[ht!]
    \centering
    \begin{tabular}{|c|c|c|c|c|c|c|}
        \hline
          molecule & \multicolumn{2}{c|}{No Pol} & \multicolumn{2}{c|}{Pol}  \\\cline{2-5}
          & Gaunt (eV) & Breit (eV) & Gaunt (eV) & Breit (eV) \\
         \hline
          CuH & 20.6381 &  -1.8419 & 20.6381 & -1.8419 \\
         \hline
          AgH &  109.7382  & -10.8607 & 109.7380 & -10.8607 \\
         \hline
          AuH &  544.0871 & -66.5148 & 544.0874 &  -66.5149  \\
         \hline
    \end{tabular}
    \caption{Energy differences between DHF and DHF-Gaunt(-Breit) level.
    }
    \label{tab:Tot_energy_RetVSPol}
\end{table}
In Table~\ref{tab:Tot_energy_RetVSPol}, the Gaunt/Breit corrections to the DHF and Pol-DHF energies are presented. These effects are quite sizable, in particular for AuH which is the system exhibiting the largest relativistic effects. In this case, the trend is monotonic and both the Gaunt and Breit effect increase going down the group of the periodic table. It is important to notice that if the absolute energy variation is taken into account, these effects are at least one order of magnitude larger than the effects generated by the quantum field. The comparison between the Breit correction in CuH (the less relativistic system) and the corresponding polaritonic energy correction, shown in Table~\ref{tab:DHF_GS_energies}, demonstrates that quite unequivocally.
These observations clearly indicate that, for a molecular system coupled to a quantum field, attention needs to be paid in omitting the Breit term. 
More striking results can be obtained if we perform the same analysis on the orbital energies. In Figure~\ref{fig:orb_mo}, the absolute value of the energy contributions (in logarithmic scale) due to the Gaunt, Breit and polaritonic terms on the energies of the occupied MOs are presented for the three systems.
\begin{figure*}[ht!]
    \centering
    \includegraphics[width=0.323\linewidth]{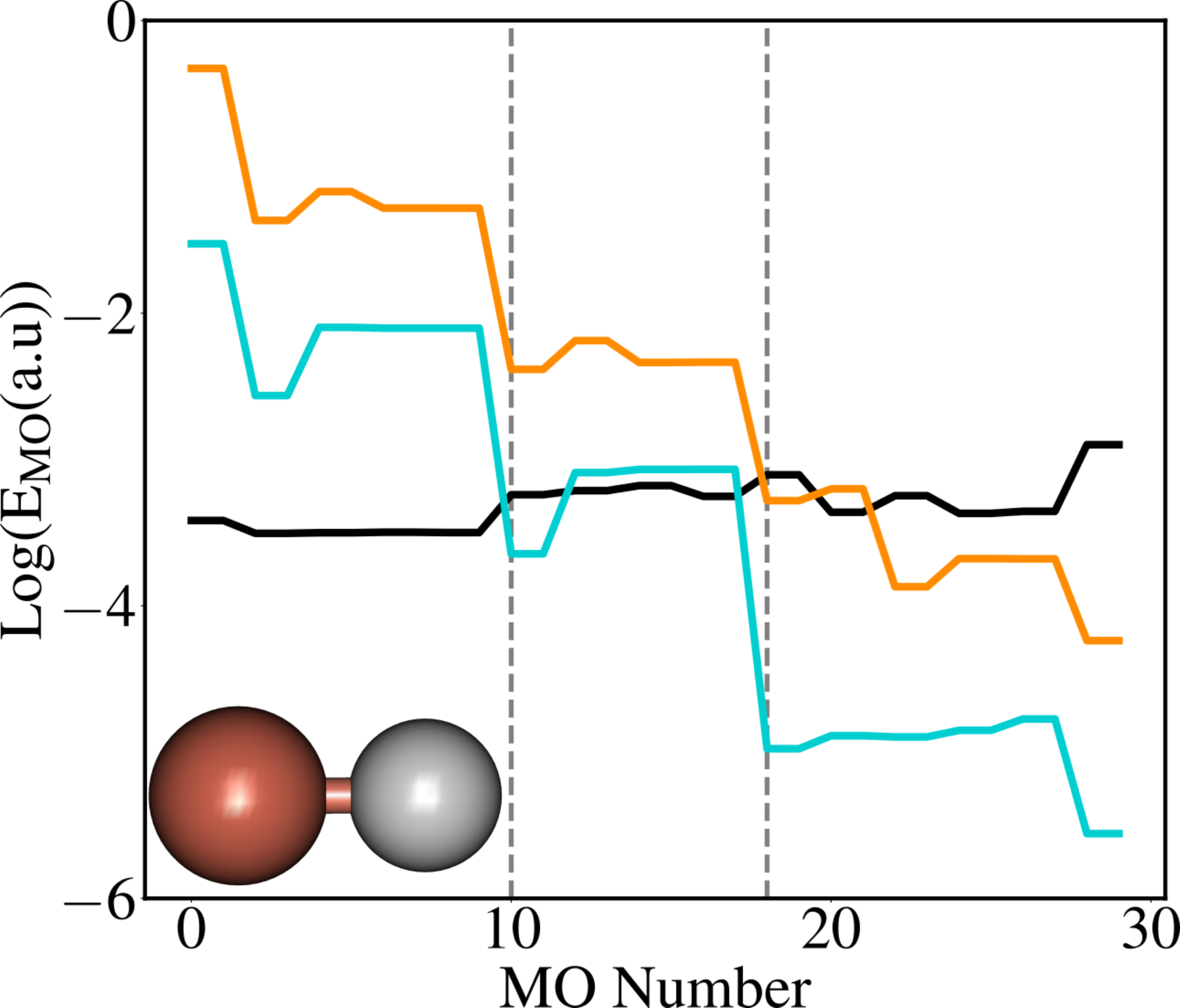}\hfill
    \includegraphics[width=0.32\linewidth]{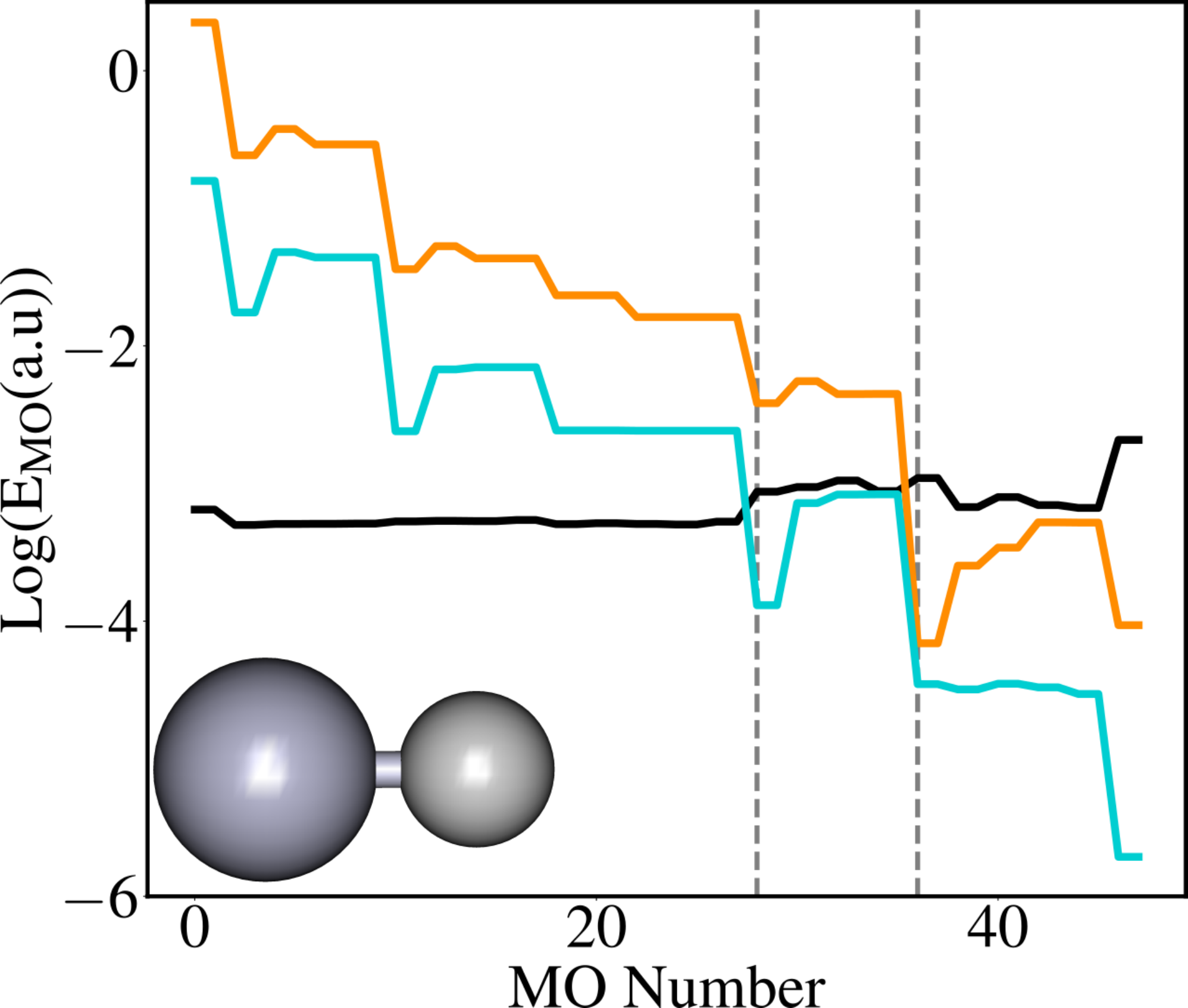}\hfill
    \includegraphics[width=0.32\linewidth]{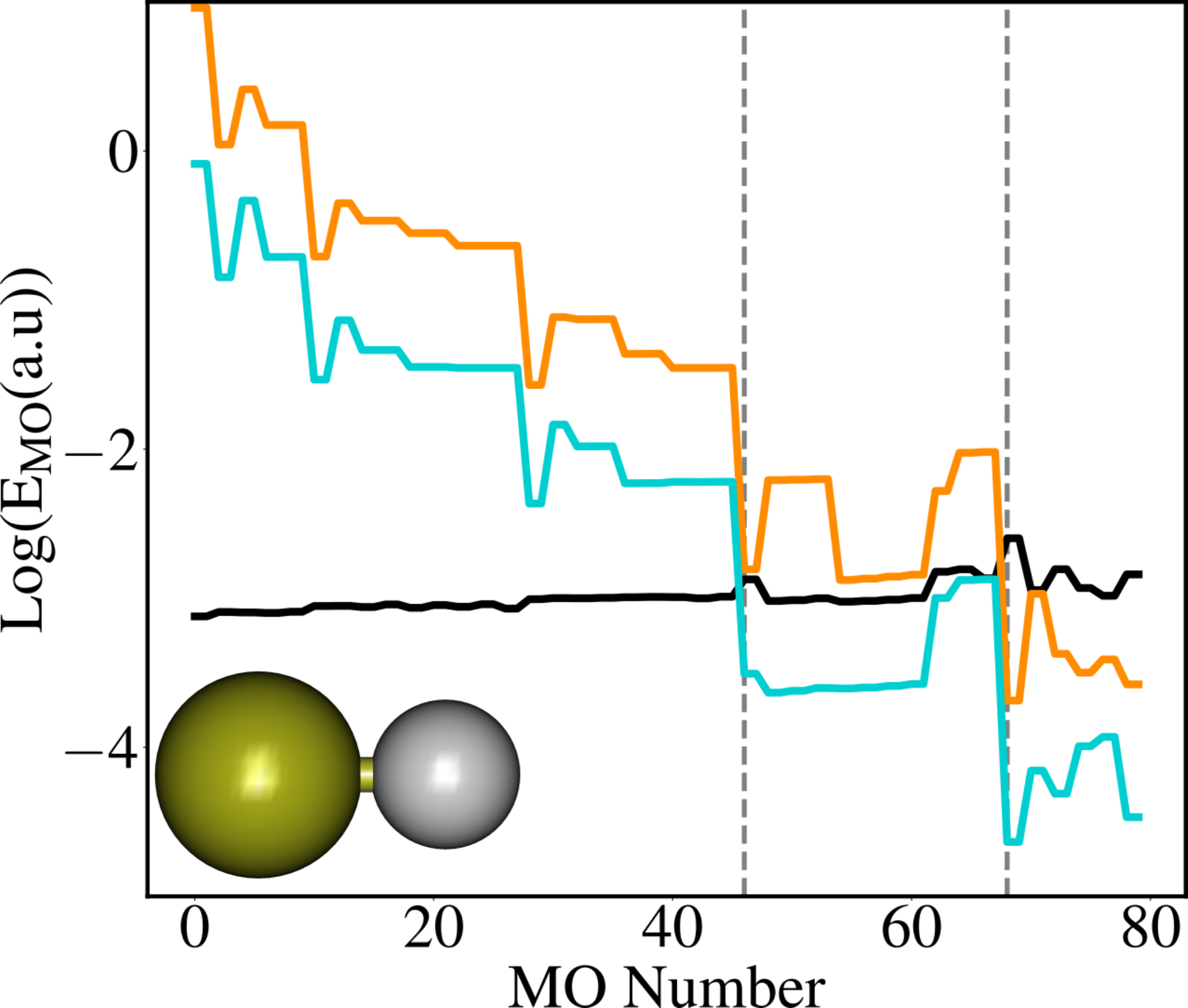}\hfill
    \caption{Absolute contribution of Polaritonic (Black), Gaunt (Orange) and Breit (Cyan) terms on the various molecular orbitals of CuH (left panel), AgH (middle panel) and AuH (right panel)). Three zones have been represented on the graph, the one on the left corresponds to the core region, the middle one is an area where Gaunt, Breit and Polaritonic contributions are comparable, and the area on the right to the valence region.}
    \label{fig:orb_mo}
\end{figure*}
The most evident aspect is the very wide variation range of the energy contributions for different molecular orbitals. In general, the effects due to the full-Breit term are much more sizable for the core orbitals, while they monotonically decrease moving toward the valence ones. This trend is clearly expected and has been documented in Ref.~\citenum{JCP_Breit_Aucar_2018}. This behavior is obviously more evident for AuH than for CuH. 
The polaritonic contribution shows instead much smaller variations, and the effect slightly increases moving toward the valence. 
In the core region, the Gaunt and Breit contributions are always larger, while in the valence region the polaritonic effects are dominant, at least, in the coupling regime used in our calculations. 
Interestingly, for the intermediate orbitals, all the effects are quite comparable, confirming that the inclusion of Breit term could induce crucial effects in molecular systems coupled to photons. For example, they might be important if the Rabi-Spittings for excitations from a core-orbital to a valence orbital are being studied, or when the induced effect of Gaunt/Breit/Polaritonic on IP and EA~\cite{PastekaPRLPhysRevLett118,DePrince_JCP_2021_QED_CC_IP_EA, JCP_Rosario_2022_IP} are investigated. However, bare in mind that such a context goes beyond the dipole-approximation and the length gauge. However, other strategies have been proposed to circumvent such an issue (see Ref.s~\citenum{PhysRevX_Rosario_2023, Castagnola2024_linear_response}).
Still, there are cases where neglecting such effects, as usually done in relativistic quantum chemistry, is still reasonable. For instance, if we are interested in excited states properties (i.e. Rabi Splitting, etc.) involving only valence electrons, a phenomenon recently analyzed by Konecny et al. in Ref.~\citenum{RelQEDDFT}, then omitting current-current interaction and the Coulomb gauge correction in the treatment is clearly meaningful. For this reason, in the next section, excited states will be investigated without including the full-Breit correction.
However, if excitations from inner orbitals (i.e. core excitations, etc.) need to be analyzed, the inclusion of the full-Breit term will be necessary.
Bear in mind that the implications of the effects presented in this section are, in some sense, minimized by the absence of electron-electron and electron-photon correlation in the treatment. For instance, the HF approximation removes the frequency dependence of the ground state energy from the frequency of the field. 
This dependence can be recovered only by including correlation into the model~\cite{nature_riso_2022}. We expect, that the inclusion of frequency dependent terms, will be crucial in particular to describe resonant processes. The inclusion of electron-electron and electron-photon correlation will be the main topic of a future follow-up paper.
\subsection{Excited state properties}
\begin{figure}[ht!]
    \includegraphics[width=\linewidth]{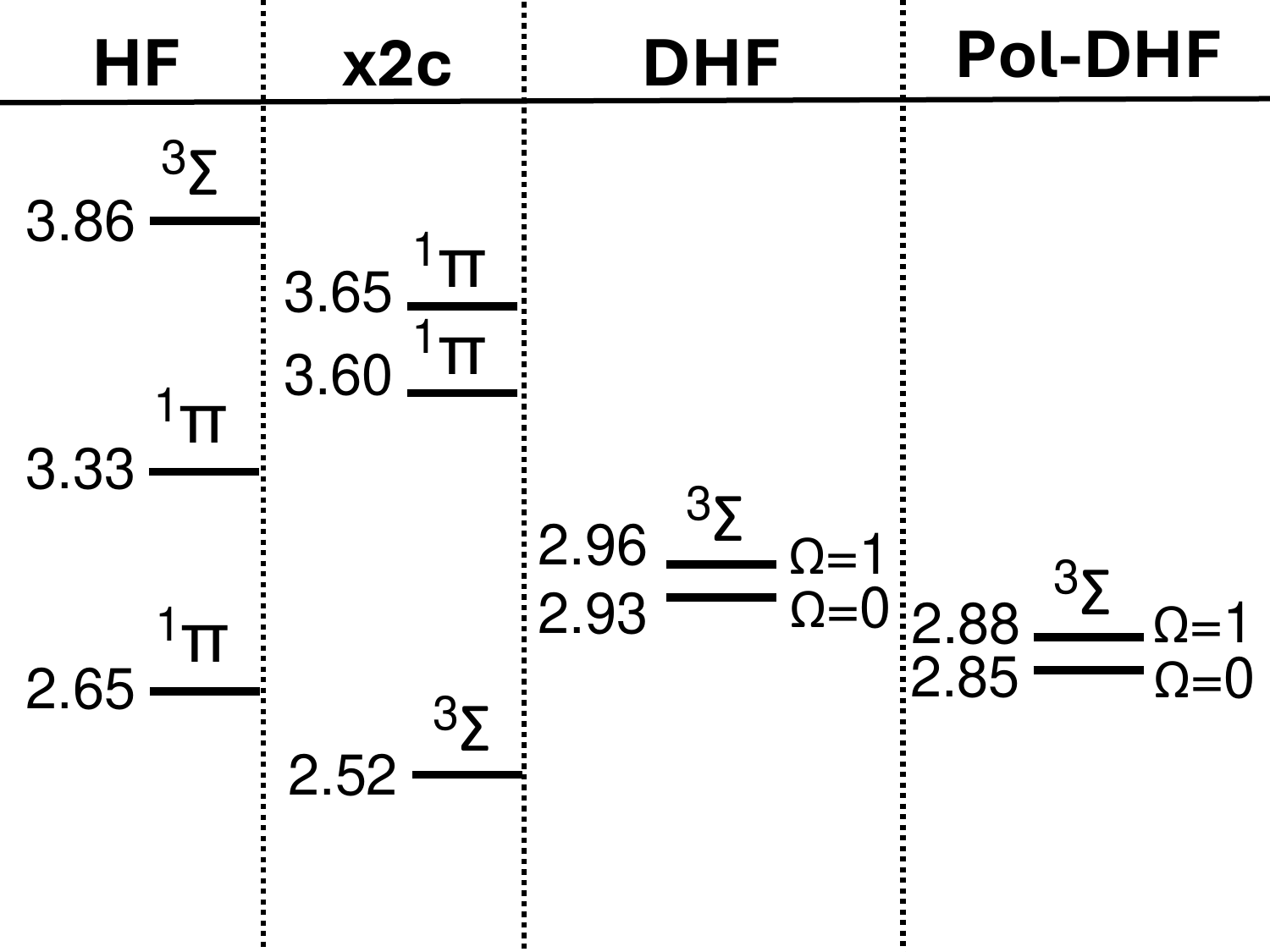}\hfill
    \caption{off-resonance $\omega=0$) excitation energies for AuH in eV with different levels of calculation.}
    \label{fig:RHF_x2c_DHF}
\end{figure}

\begin{figure}
    \centering
    \includegraphics[width=1.0\linewidth]{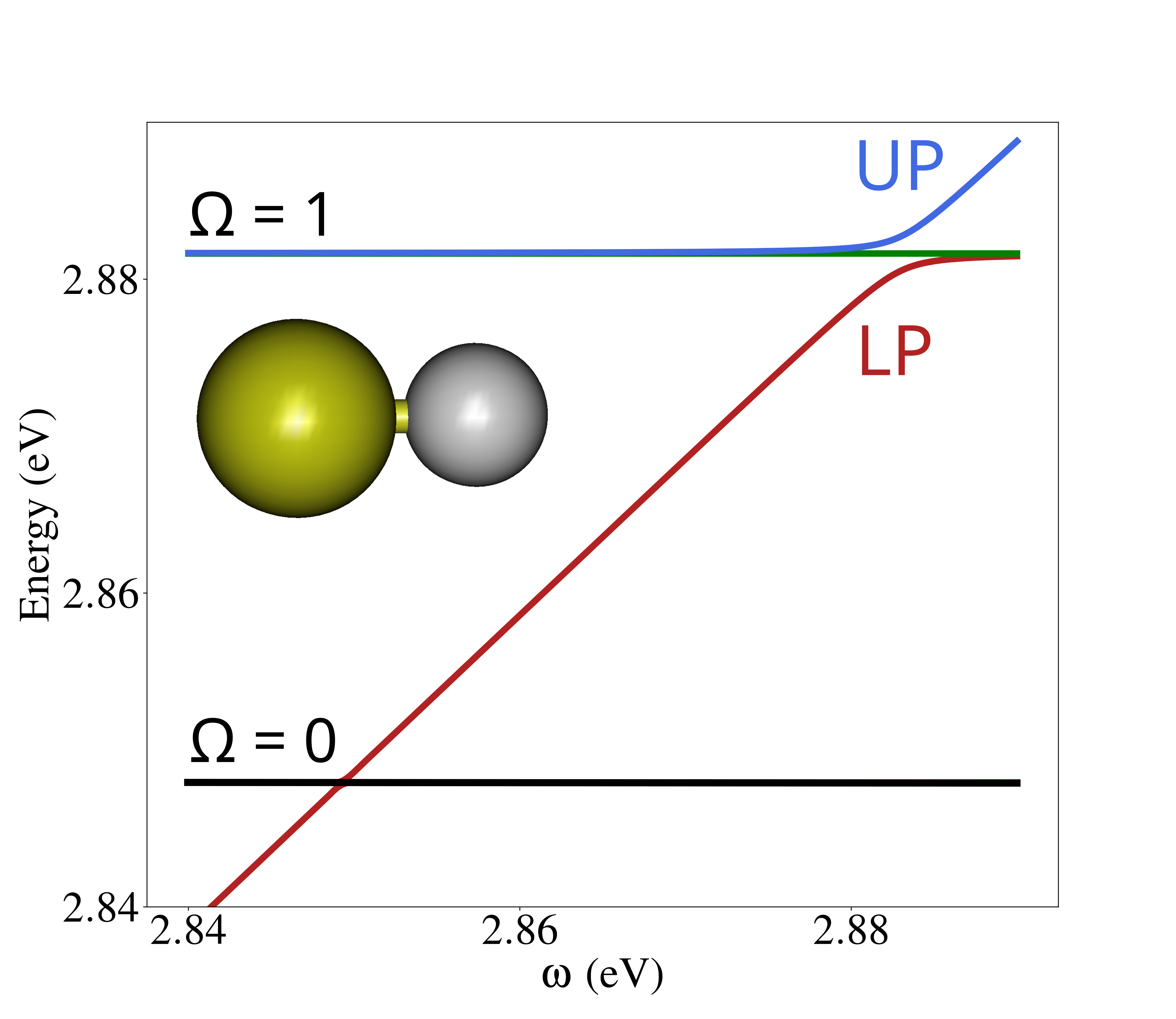}\hfill
    \caption{Excitation energies evaluated with linear response Pol-DHF as a function of the cavity frequency for AuH. UP stands for Upper Polariton and LP for Lower Polariton.}
    \label{fig:triple_rabi}
\end{figure}

In the previous section, we focused on the field induced effects on the ground state properties of three metal hydrides of the 11$^{\mathrm{th}}$ group of the periodic table.
In this section, we analyze in details the effects generated by the photons on the optical properties of the AuH complex~\cite{AuHExp1964,AuHTheo2000}. Similar analysis for CuH and AgH is  presented in section 2 of the SI. 
In this system, because of strong relativistic effects, we expect to observe spectra very different compared to those simulated without taking relativity into account.
This is obvious looking at Figure~\ref{fig:RHF_x2c_DHF}.
As it can be seen, at the HF level, the triplet state ($^{3}\Sigma$) is found to be higher in energy than the $\Pi$ singlets states, in disagreement with the experimental data from Ref.s~\citenum{AuHExp1964,AuHTheo2000}. The inclusion of a spin free relativistic correction partially resolve this issue, but it underestimates the energy by about $\sim$ 0.45 eV compared to DHF calculations. DHF does not only reproduce the right ordering of the states, but it also describes the breaking of the degeneracy to form (from the triplets) the $\Omega=0$ and $\Omega=1$ states. For an improved readability of the plot Figure~\ref{fig:RHF_x2c_DHF} the Pol-DHF data have been calculated off-resonance. This choice highlights the energy shift due to the dipole self energy. In resonance conditions, the Rabi splitting (discussed later in Fig.~\ref{fig:triple_rabi}) would also be visible. 
Comparing this data with the bare electronic DHF ones, we see that the field (in on resonance conditions) induces a stabilization of the $\Omega=0$ and $\Omega=1$ states without inducing any change in the ordering of the states.\\
Finally, in Fig.\ref{fig:triple_rabi} we have reported the dispersion of the AuH excitation energies as a function of the cavity frequency. The excitations falling in the investigated energy range, refers to the $\Omega = 0 $ and $\Omega = 1$ states obtained by the splitting of the triplets. For a non-relativistic system we should not observe any Rabi splitting due to the $\Delta J = \pm 1$ selection rule. In this case instead, due to the strong spin-orbit coupling, a sizable splitting can be observed at the crossing between the first photonic replica of the ground state ($^{1}\Sigma$) with the $\Omega = 1$ state. For this system, the energy difference between $\Omega = 0 $ and $\Omega = 1 $, associated to spin-orbit coupling, is one order of magnitude larger (0.03 eV) than the Rabi-splitting ($\sim 10^{-3}$ eV). However, a word of caution is necessary. In this work, for computational reasons, we have used a contracted basis set which is usually not  recommended when working in 4-component frameworks. In order to assess the important of such a choice, we have performed calculations with uncontracted basis sets for all the systems at the ground state levels, and for the excited state of CuH. At the ground state level the results are not significantly affected. For the excited state calculations, for CuH, the use of an uncontracted basis set reduces the magnitude of the Rabi-Splitting, rendering the latter of comparable size compared to the spin-orbit coupling. This is rather reasonable since the transition is \textit{a priori} forbidden and a weak spin-orbit coupling induces a weaker transition dipole moment. Still, even for a system where the spin-orbit coupling is not strong, it appears that polaritonic effects can be of a comparable size (more details are provided in section 3 of the SI).
This spin-orbit coupling induced singlet-triplet Rabi-splitting was already reported for a different system by Konecny et al. in Ref.~\citenum{RelQEDDFT}, though the systems are different, no qualitative differences in behavior are expected. These observations clearly demonstrate how the electromagnetic field can be used to manipulate and control inter-system crossing processes and consequently the phosphorescence of complexes containing heavy atoms. In section 2 of the SI, a similar discussion is also reported for CuH and AgH.

\section{Conclusions}

In this paper, we proposed a reformulation of relativistic QED allowing for an easier development of \abinit methodologies to simulate heavy atoms molecular complexes in strong coupling conditions.
Using this theoretical ground, we reported the development and implementation of the first relativistic polaritonic wave function based \abinit method, namely Pol-DHF. The theory has been presented starting from the standard Lagrangian and has been derived into a usable implementation of the Pol-DHF code. Considering the possible competition with the polaritonic effects, the importance of radiative QED corrections have been addressed though their inclusion is left for future work. After providing a road-map to the implementation, we presented applications of Pol-DHF to three metal hydrides: CuH, AgH and, AuH. These systems were an excellent test case in order to assess the magnitude of the polaritonic effects in comparison to the relativistic effects. To do so, we evaluated the influence of the polaritonic effects on the $nd \rightarrow (n+1)s\sigma$ gap for the three systems. The polaritonic effects resulted having the largest relative influence on AuH for which relativistic effects are more prominent. 
Afterward, we provided a detailed analysis of the competition between the Gaunt, Breit and polaritonic effects on the ground state and orbital energies. 
We have verified that, in the polaritonic context, even though the full-Breit term represents a significantly larger contribution to the total ground state energy, its effect remained extremely orbital dependent. 
In particular, its impact is much stronger on the core orbitals compared to the valence, where instead the polaritonic effects dominate in particular if very strong coupling values can be reached. Therefore, we could conclude that neglecting the full-Breit term in strong coupling conditions is not always possible, in particular, it can be done only if properties involving exclusively valence orbitals need to be investigated. A similar study could be conducted adding effective potentials, or using BSQED techniques, to include radiative QED corrections and comparing the magnitude of their effects to the polaritonic ones, this will surely be the topic of a future work. \\  
Lastly, we presented excited state calculations for AuH at the TD-Pol-DHF level. As already observed by Konecny et al.~\cite{RelQEDDFT}, we have shown that using a fully relativistic polaritonic theory, the appearance of Rabi-Splittings at the crossing between singlet and triplet potential energy surfaces can be observed. Furthermore, we have provided a comparison between polaritonic and relativistic effects (e.g. Rabi splitting vs spin orbit coupling). Lastly, Pol-DHF represents the perfect platform upon which developing more elaborate methods (e.g including dynamic and static correlation). Pol-DHF also opens the door for the development of polaritonic x2c methods and would represent a point of reference that would facilitate such endeavor.
We strongly believe that the methodologies and the applications presented in this paper can represent a significant step toward the simulation of relativistic molecular systems strongly coupled to photons, field that is recently finding many interesting applications in photochemistry but also in spintronics and quantum computation. 

\section{Data Availability Statement}
The development version of PySCF used to produce the datasets presented in this manuscript together with the inputs files and the geometries are available at the Zenodo link~\cite{zenodo}. 
\section{Supporting Information}
Additional theoretical details regarding the derivation of the QED Hamiltonian and the length gauge transformation. Excitation energies with respect to the cavity frequency for CuH and AgH. Ground state total energies at the DHF and Pol-DHF level for CuH using an uncontracted basis set. Energy differences between DHF and DHF-Gaunt(-Breit) level for CuH with an uncontracted basis set. Gaunt, Breit and Polaritonic contribution in absolute value to the total energy with respect to MO number for CuH using an uncontracted basis set. Excitation energies with respect to the cavity frequency for CuH using an uncontracted basis set.

\section{Acknowledgement}
The authors acknowledge Michael Ruggenthaler, Lukas Konecny, Rosario Roberto Riso and Giovanni Bistoni for insightful discussions.
G. T., R. R. and E. R. acknowledge funding from the European Research Council (ERC) under the European Union’s Horizon Europe Research and Innovation Program (Grant ERC-StG-2021-101040197-QED-SPIN). H. K. acknowledges funding from the European Research Council (ERC) under the European Union’s Horizon 2020 Research and Innovation Program (Grant Agreement 101020016).

%
%
%

\title{Supporting Information to: ``A comprehensive theory for relativistic polaritonic chemistry: a four components \abinit treatment of molecular systems coupled to quantum fields''
}

\author{Guillaume Thiam}
\email{guillaume.thiam@unipg.it}
\author{Riccardo Rossi}%
\affiliation{Dipartimento di Chimica, Biologia e Biotecnologie, Universit\`a degli Studi di Perugia, Via Elce di Sotto, 8,06123, Perugia, Italy}

\author{Henrik Koch}
\affiliation{Department of Chemistry, Norwegian University of Science and Technology, 7491 Trondheim, Norway}

\author{Leonardo Belpassi}
\affiliation{Istituto di Scienze e Tecnologie Chimiche “Giulio Natta” del CNR (CNR-SCITEC), Via Elce di Sotto, 8, 06123 Perugia, Italy}

\author{Enrico Ronca}
\email{enrico.ronca@unipg.it}
\affiliation{Dipartimento di Chimica, Biologia e Biotecnologie, Universit\`a degli Studi di Perugia, Via Elce di Sotto, 8,06123, Perugia, Italy}

\maketitle

\noindent \textbf{Supporting information:}\\
\textbf{Section 1:} Additional theoretical details. \\
\textbf{Section 2:} Excited States properties of CuH and AgH. \\
\textbf{Figure SI1:} Excitation energies with respect to the cavity frequency for CuH.\\
\textbf{Figure SI2:} Excitation energies with respect to the cavity frequency for AgH.\\
\textbf{Section 3:} Effect of using an uncontracted basis set. \\
\textbf{Table SI1:} Ground state total energies at the DHF and Pol-DHF level for CuH using an uncontracted basis set. \\
\textbf{Table SI2:} Energy differences between DHF and DHF-Gaunt(-Breit) level for CuH with an uncontracted basis set\\
\textbf{Figure SI3:} Gaunt, Breit and Polaritonic contribution in absolute value to the total energy with respect to MO number for CuH using an uncontracted basis set.\\
\textbf{Figure SI4:} Excitation energies with respect to the cavity frequency for CuH using an uncontracted basis set.\\

\section{Theory derivation}

In this section, we follow the formal derivation of relativistic QED theory usually presented in physics text books\cite{CohenTanoudjiQED, ReiherBook} to develop a Hamiltonian formalism that can be applied to formulate new \abinit methodologies for the simulation of polaritonic molecular systems. This choice has been meant to render the overall discussion accessible to a broad chemistry audience. The proposed methodology is then used  to develop the first Hartree-Fock (HF) based approach for relativistic molecular systems strongly coupled to quantum fields.  
For convenience reasons, Gaussian units will be used during the whole derivation unless specified otherwise. 
\subsection{\label{sec:level2.1} The Quantum Electrodynamics Lagrangian}

We start our derivation from the definition of a Lagrangian describing at the same time the relativistic molecular system, the electromagnetic field and their interaction. In the following, all Greek letters indices span the components of 4 indices vectors (from 0 to 3), whereas latin letters only span the spatial components of the vector (from 1 to 3). Moreover, Einstein summation conventions are used.

To describe matter, we use the standard Lagrangian density for Dirac fields:
\begin{equation}
    \mathcal{L}_{\mathrm{Dirac}}= \bar{\Psi}_{e}(i\hbar c\gamma^{\mu}\partial_{\mu}-m_{e}c^2)\Psi_{e}
\end{equation}
where $c$ is the speed of light and $m_e$ is the mass of the electron. 
The matrices $\gamma^{\mu}$ are defined as:
\begin{equation}
\gamma^{0}= \begin{pmatrix}
\mathbf{1} & \mathbf{0} \\ \mathbf{0} & -\mathbf{1}
\end{pmatrix}  \hspace{5mm}
\gamma^{i}= \begin{pmatrix}
\mathbf{0} & \sigma^{i} \\ -\sigma^{i} & \mathbf{0}
\end{pmatrix}
\end{equation}
with $\boldsymbol{\sigma}^i$ representing the Pauli matrices:
\begin{equation}
\sigma^{x}= \begin{pmatrix}
0 & 1 \\ 1 & 0
\end{pmatrix}  \hspace{5mm}
\sigma^{y}= \begin{pmatrix}
0 & -i \\ i & 0
\end{pmatrix} \hspace{5mm}
\sigma^{z}= \begin{pmatrix}
1 & 0 \\ 0 & -1    
\end{pmatrix}
\end{equation}
The $\gamma^{\mu}$ matrices are needed to construct Lorentz invariant quantities and naturally include the spin-orbit coupling in the theory.
$\partial_{\mu}=\Big(\partial_t,\partial_x,\partial_y, \partial_z\Big)$ is the 4-derivative and $\Psi_{e}$ are the electron spinor fields having $\bar{\Psi}_{e}=\Psi_{e}^{\dagger}\gamma^0$ as adjoint.

The dynamics of the electromagnetic field is described instead by the Lagrangian density ($\mathcal{L}_{\mathrm{Maxwell}}$):
\begin{equation}\label{eq:LagrangianMaxwell}
    \mathcal{L}_{\mathrm{Maxwell}}=-\frac{1}{16\pi}F^{\mu\nu}F_{\mu\nu}
\end{equation}
where the field tensor $F^{\mu\nu}$:
\begin{equation}\label{eq:field_tensor}
F^{\mu\nu}=\partial^{\mu}A^{\nu}-\partial^{\nu}A^{\mu},   
\end{equation}
depends on the 4-vector potential $A^{\nu} = (\phi, \mathbf{A})$
and mediates all the electromagnetic interactions. In the context of polaritonic chemistry, such a term accounts both for the field induced by the electron and the one inherent to the confinement of the molecular system in the cavity. Maxwell's equations allow to define auxiliary scalar and vector potentials, respectively $\phi$ and $\mathbf{A}$. 
These potentials are not uniquely defined, and many potentials lead to the same electric and magnetic field. This is referred to as gauge freedom~\cite{CohenTanoudjiQED, Jackson2002Gauge}: 
\begin{align}
    & \mathbf{A}'=\mathbf{A} + \nabla f \\
    & \phi'=\phi - \frac{1}{c}\frac{\partial f}{\partial t}
\end{align}
where $f$ is a scalar function.
$\phi$ is related to the electrostatic component of the electric field, and, in Coulomb gauge, to the longitudinal part of the electric field. On the other hand, $\mathbf{A}$ is related to the magnetic field, and, in Coulomb gauge, to the transverse component of the electric field. We remind the reader that the electric field can be expressed in terms of scalar and vector potentials $\mathbf{E} = -\mathbf{\nabla}\phi-\frac{1}{c}\partial_{t}\mathbf{A}$ and the magnetic field in terms of vector potential $\mathbf{B} = \nabla \times \mathbf{A}$. 
The choice of the Lagrangian density for the electromagnetic field is not unique, and different equivalent forms can be used depending on the gauge. Lagrangian~\ref{eq:LagrangianMaxwell} is usually the most convenient choice in Coulomb gauge ($\nabla\cdot \mathbf{A} = 0$~\cite{Jackson2002Gauge,CoulombGaugeM_Stewart_2003}). This Lagrangian is not always convenient if other gauges~\cite{peskin1995introduction,QFTongCambridge} (e.g. Lorenz gauge\cite{Jackson2002Gauge}) need to be used.  
The light and matter terms are coupled via interaction contributions:
\begin{equation}\label{eq:IntLagrangian}
    \mathcal{L}_{\mathrm{Int}} = -\frac{1}{c}j_{\mu}A^{\mu} -\frac{1}{c}j_{\mu}A_{\text{ext}}^{\mu}
\end{equation}
where $A_{\text{ext}}^{\mu}$ is an external four vector potential that can be associated to the nuclei, to an external (non-dynamical) electromagnetic field, etc. At the moment, its definition remain general and will be specified when needed. 

The 4-current $j^{\mu}= (c\rho,\mathbf{j})$ in Eq.~\ref{eq:IntLagrangian} can be expressed as a function of $\Psi_{e}$ and $\gamma^{\mu}$ as:
\begin{equation}\label{eq:current}
    j^{\mu} = ec\bar{\Psi}_{e}\gamma^{\mu}\Psi_{e}.
\end{equation}

 Finally, the complete QED Lagrangian takes the form:
 \begin{equation}
    \mathcal{L}_{\mathrm{QED}}=\mathcal{L}_{\mathrm{Dirac}}+\mathcal{L}_{\mathrm{Maxwell}}+\mathcal{L}_{\mathrm{Int}}\label{eq:Lagrangian_RQC}
\end{equation}   
Note that only the fermionic ($\Psi_{e}$) and electromagnetic fields ($A^{\mu}$) will be treated as dynamical variables~\cite{ReiherBook}. 
Eq.~\ref{eq:Lagrangian_RQC} will be the starting point for the development of the Hamiltonian formalism derived in the following.

\subsection{\label{sec:level2.2} Hamiltonian formulation in Coulomb gauge}

Starting from Lagrangian~\ref{eq:Lagrangian_RQC} a Hamiltonian formulation of the theory can be derived by performing a Legendre transform:
\begin{equation}\label{eq:ham_density}
    \mathcal{H}_{\mathrm{QED}}=\Pi_\mu \Dot{A}^\mu + \pi\Dot{\Psi}_e-\mathcal{L}_{\mathrm{QED}}
\end{equation}
where the conjugate momenta are given by:
\begin{equation}\label{eq:conj_momenta}
    \pi=\frac{\partial\mathcal{L}_{\mathrm{QED}}}{\partial\dot{\Psi_e}} \qquad
    \Pi_{\mu} = \frac{\partial\mathcal{L}_{\mathrm{QED}}}{\partial\dot{A^{\mu}}}
\end{equation}

Substituting Eq.~\ref{eq:Lagrangian_RQC} in Eq.~\ref{eq:ham_density} and using the Green theorem: 
\begin{equation}
\int_V \nabla\phi\cdot \nabla\phi d\mathbf{r} = \int_\Sigma \phi \nabla \phi \cdot d\boldsymbol{\sigma} - \int_V \phi \nabla^2 \phi d\mathbf{r}
\end{equation}
where the surface integral is  zero, we obtain the following Hamiltonian:
\begin{widetext}
 \begin{align}\label{eq:H}
H & = -\frac{1}{8\pi}\int \left[\underbrace{-\phi \nabla^2 \phi}_{\mathbf{E}^{2}_{\text{long}}} \underbrace{-\frac{\mathbf{\dot{A}}\cdot \mathbf{\dot{A}}}{c^2} - (\nabla\times\mathbf{A})\cdot(\nabla\times\mathbf{A})}_{\mathbf{E}_{\mathrm{trans.}}^{2} + \mathbf{B}^{2}}\right] d\mathbf{r} \nonumber \\
& + \int \Psi_{e}^{\dagger}[c\alpha_{i} (-i\hbar \nabla_i - \frac{e}{c} A_{i} - \frac{e}{c} A^{\text{ext}}_{i} ) +\beta m_{e}c^2 )] \Psi_{e} d\mathbf{r}  + \int \phi \rho d\mathbf{r} + \int \phi_{ext} \rho d\mathbf{r} 
\end{align}   
\end{widetext}

\subsubsection{Relativistic Pauli-Fierz Hamiltonian in the length gauge}\label{subsec:RPF}
When investigating molecular systems, it is usually more convenient to apply a unitary transformation to the field modes, allowing for a direct coupling between the field operators and the molecular dipole. This transformation is known as the length-gauge transformation:
\begin{equation}\label{eq:length_gauge_trans}
    U = \exp(\frac{ie}{\hbar c} \mathbf{A}_{\text{dm}}(\vec{0})\cdot \mathbf{R}) 
\end{equation}
where $\mathbf{R} = \I \Psi^{\dagger} \mathbf{r} \Psi$. Hamiltonian 26 in the main text can be transformed in the length-gauge form by application of Eq.~\ref{eq:length_gauge_trans}:
\begin{equation}\label{eq:transfo}
 H^{l}_{RPF} = U^{\dagger} H_{QED} U
\end{equation}
followed by a rotation of the photonic coordinates associated with the mode \textbf{k}$_{\text{dm}}$: $\Tilde{U} = \exp(-i\frac{\pi}{2} \sum_{\vec{\epsilon}} a'^{\dagger}_{\text{dm},\vec{\epsilon}} a'_{\text{dm},\vec{\epsilon}})$ which ensures the photonic part to be real. Transformation in Eq.\ref{eq:transfo} induces a cancellation of the $\boldsymbol{\alpha} \cdot \mathbf{A}_{\text{dm}}$ term due to the change in the momentum. The light-matter coupling is now related to the molecular dipole.
\subsection{Length gauge transformation}\label{app:lenghtgauge}
In the article, we presented the results of the application of the length gauge transformation. In this appendix, we provide a detailed derivation of such results.
Since most of the terms of Hamiltonian 24 in the main text commute with the $U$ operator, only two terms get modified, the one involving the momentum $p$ and the one involving the photon number operator $ a^{\dagger}_{\tau}a_{\tau}$. Using the special case of the Becker-Campbell-Hausdorff formula:
\begin{align}
    & \I \Psi^{\dagger} c\alpha^{i}p_{i}\Psi \nonumber \\
    & \longrightarrow \I \Psi^{\dagger} c\alpha^{i}p_{i}\Psi - \left[\frac{ie}{\hbar c} \mathbf{A}_{\text{dm}}(\vec{0})\cdot \mathbf{R}, \I \Psi^{\dagger} c\alpha^{i}p_{i} \Psi \right].
\end{align}
Let us consider the commutator only:
\begin{align}
    & \frac{ie}{\hbar c} A_{j, \text{dm}} \left[\int d^{3}r'\Psi^{\dagger}(\vec{r}') r^{j} \Psi(\vec{r}'), \I \Psi^{\dagger}(\vec{r}) c\alpha^{i} p_{i} \Psi(\vec{r})  \right] \\
    & = \frac{ie}{\hbar c} A_{j} \int d^{3}r'\Psi^{\dagger}(\vec{r}') \Psi(\vec{r}') \underbrace{\left[r^{j}, p_{i} \right]}_{= i\hbar \delta^{j}_{i}} \I \Psi^{\dagger}(\vec{r}) c\alpha^{i} \Psi(\vec{r}) \\
    & = -e A_{j, \text{dm}}  \underbrace{\int d^{3}r'\Psi^{\dagger}(\vec{r}') \Psi(\vec{r}')}_{1} \I \Psi^{\dagger}(\vec{r}) \alpha^{j} \Psi(\vec{r})  \\
    & = -  \I \Psi^{\dagger}(\vec{r}) e A_{j, \text{dm}}  \alpha^{j} \Psi(\vec{r}) 
\end{align}
and therefore
\begin{equation}
    \I \Psi^{\dagger} c\alpha^{i}p_{i}\Psi \longrightarrow \I \Psi^{\dagger} c\alpha^{i}\left( p_{i} + \frac{e}{c} A_{i, \text{dm}} (\vec{0})\right) \Psi.
\end{equation}
It is important to emphasize that the dipole approximation strongly simplifies the expression of the terms involving the momentum. The other noticeable fact is that the terms involving $\alpha^{i} A_{i}(\vec{0})$ cancels out (the translated momentum bring out a $+e\mathbf{A}_{\text{dm}}$ term canceling out with the unmodified $-e\mathbf{A}_{\text{dm}}$ term). \\
Let us now look at the modification of the photon number operator. The Becker-Campbell-Hausdorff formula implies that:
\begin{equation}
    a'^{\dagger}_{\text{dm},\vec{\epsilon}}a'_{\text{dm},\vec{\epsilon}} \longrightarrow a'^{\dagger}_{\text{dm},\vec{\epsilon}}a'_{\text{dm},\vec{\epsilon}} -\left[ \frac{ie}{\hbar c}\mathbf{A}_{\text{dm}} (\vec{0})\cdot \mathbf{R},a'^{\dagger}_{\text{dm},\vec{\epsilon}}a'_{\text{dm},\vec{\epsilon}} \right] + ...
\end{equation}
The first commutator gives:
\begin{equation}
    -\frac{ie}{\hbar c} \frac{C \mathbf{R} \cdot \vec{\epsilon} }{\sqrt{\omega_{\text{dm}}}}\left[a'_{\text{dm},\vec{\epsilon}} - a'^{\dagger}_{\text{dm},\vec{\epsilon}} \right]
\end{equation}
where $C = \sqrt{\frac{2\pi\hbar}{V}}$. The second commutator gives:
\begin{equation}
    + \left(\frac{Ce\mathbf{R}\cdot \vec{\epsilon}}{\hbar c} \right)^{2}\frac{1}{\omega_{\text{dm}}}.
\end{equation}
Therefore, the Hamiltonian then reads:
\begin{align}
   & \mathrm{H}^{l}_{RPF} = \nonumber \\
    &\I \Psi^{\dagger} \{c \alpha^{i} \left(p_{i}  -\frac{e}{c}A_{\text{om},i}(\mathbf{r}) -\frac{e}{c}A_{\text{ext},i}(\mathbf{r})\right) + \beta m_{e} c^{2}\} \Psi \nonumber \\
    & + \frac{1}{2}\I d\mathbf{r}' \Psi^{\dagger}(\mathbf{r}) \Psi(\mathbf{r}) \frac{1}{\lvert \mathbf{r} - \mathbf{r}^{'} \rvert}\Psi^{\dagger}(\mathbf{r}') \Psi(\mathbf{r}') \nonumber \\
    & + \sum_{\vec{\epsilon}} \hbar \omega_{\text{dm}} \{ a'^{\dagger}_{\text{dm},\vec{\epsilon}}a'_{\text{dm},\vec{\epsilon}}-\frac{ie}{\hbar c} \frac{C \mathbf{R} \cdot \vec{\epsilon} }{\sqrt{\omega_{\text{dm}}}}\left[a'_{\text{dm},\vec{\epsilon}} - a'^{\dagger}_{\text{dm},\vec{\epsilon}} \right] \nonumber \\  
    & + \left(\frac{Ce\mathbf{R}\cdot \vec{\epsilon}}{\hbar c} \right)^{2}\frac{1}{\omega_{\text{dm}}}+\frac{1}{2}\} \nonumber \\
    & + \sum_{\tau} \hbar \omega_{\tau} \left( a'^{\dagger}_{\tau}a'_{\tau}+ \frac{1}{2}\right).
\end{align}
The final Hamiltonian in length gauge appears as:
\begin{align}\label{eq:H_RPF_length_final}
    \mathrm{H}^{l}_{RPF} &= \I \Psi^{\dagger} h \Psi \nonumber\\ 
    &+ \frac{1}{2}\I d\mathbf{r}' \Psi^{\dagger}(\mathbf{r}) \Psi(\mathbf{r}) \frac{1}{\lvert \mathbf{r} - \mathbf{r}^{'} \rvert} \Psi^{\dagger}(\mathbf{r}') \Psi(\mathbf{r}') \nonumber \\
    & + \sum_{\vec{\epsilon}} \hbar \omega_{\text{dm}} \Bigg[ a_{\text{dm},\vec{\epsilon} }^{\dagger}a_{\text{dm},\vec{\epsilon}}-\frac{e}{\hbar c} \frac{C \mathbf{R}  \cdot \epsilon_{\tau} }{\sqrt{\omega_{\text{dm}}}}(a_{\text{dm},\vec{\epsilon}} + a_{\text{dm},\vec{\epsilon}}^{\dagger} ) \nonumber \\    
    &+ \left(\frac{Ce\mathbf{R} \cdot \epsilon_{\tau}}{\hbar c} \right)^{2}\frac{1}{\omega_{\text{dm}}}+\frac{1}{2}\Bigg] 
\end{align}
where $h$ is :
\begin{equation}\label{eq:def_h}
   h = c \alpha^{i} \left( p_{i} -\frac{e}{c}A_{\text{om},i}(\mathbf{r}) -\frac{e}{c}A_{\text{ext},i} \right) + \phi_{\text{ext}} + \beta m_{e} c^{2}
\end{equation}

where $C= \sqrt{\frac{2\pi\hbar}{V}}$. Here, $a_{\text{dm},\vec{\epsilon}}$, $a_{\text{dm},\vec{\epsilon}}^{\dagger}$ have been relabeled to emphasize that, due to the length gauge transformation, the latter are modified. In fact, $\langle a^{\dagger}_{\text{dm},\vec{\epsilon}}a_{\text{dm},\vec{\epsilon}} \rangle$ does not coincide with the number of photons related to the mode \textbf{k}$_{\text{dm}}$ anymore~\cite{Flick_QED_DFT_ACS_2018}.
This Hamiltonian is the analogous of the standard Pauli-Fierz Hamiltonian usually applied in non-relativistic polaritonic chemistry~\cite{Flick_PRA_QED_DFT_2014, PhysRevB_2019_QED_Ronca, phys_rev_x_Ronca_2020}, where molecular orbitals are replaced by molecular spinors. Equation~\ref{eq:H_RPF_length_final} has an apparent origin dependence, coming from the presence of the dipole operator $\mathbf{R}$. This problem can be solved by the coherent state transformation $U_c = \Pi_{\vec{\epsilon}}\exp(z_{\vec{\epsilon}}a^{\dagger}_{\text{dm}, \vec{\epsilon}} - z^*_{\vec{\epsilon}}a_{\text{dm}, \vec{\epsilon}})$ with $z_{\vec{\epsilon}} = \frac{e\sqrt{2\pi}}{c\sqrt{\hbar \omega_{\text{dm}} V}}\langle\mathbf{R}  \cdot \vec{\epsilon}\rangle$~\cite{phys_rev_x_Ronca_2020}. Applying this transformation to the Relativistic-Pauli-Fiertz (RPF) Hamiltonian, we obtain:
\begin{align}\label{eq:H_QED_length_coherent}
    \mathrm{H}^{l}_{RPF} &= \I \Psi^{\dagger} h \Psi \nonumber \\
    &+ \frac{1}{2}\I d\mathbf{r}'\Psi^{\dagger}(\mathbf{r}) \Psi(\mathbf{r}) \frac{1}{\lvert \mathbf{r} - \mathbf{r}^{'} \rvert} \Psi^{\dagger}(\mathbf{r}') \Psi(\mathbf{r}')\nonumber \\
    & + \sum_{\vec{\epsilon}} \hbar \omega_{\text{dm}} \left( a^{\dagger}_{\vec{\epsilon}}a_{\vec{\epsilon}} +\frac{1}{2}\right) + \left(\frac{e}{c}\sqrt{\frac{2\pi}{V}}\vec{\epsilon}\cdot \left[ \mathbf{R} - \langle \mathbf{R} \rangle \right] \right)^{2} \nonumber \\
    &-\sqrt{\hbar \omega_{\text{dm}}}\left( \frac{e}{c} \sqrt{ \frac{2\pi}{V}}\vec{\epsilon}\cdot \left[ \mathbf{R} - \langle \mathbf{R} \rangle \right] \right) \left( a_{\vec{\epsilon}} + a^{\dagger}_{\vec{\epsilon}} \right),
\end{align}

\section{Excited states properties of CuH and AgH}\label{app:RabiCuHAgH}

\begin{figure}[h]
    \centering
    \includegraphics[width=1.0\linewidth]{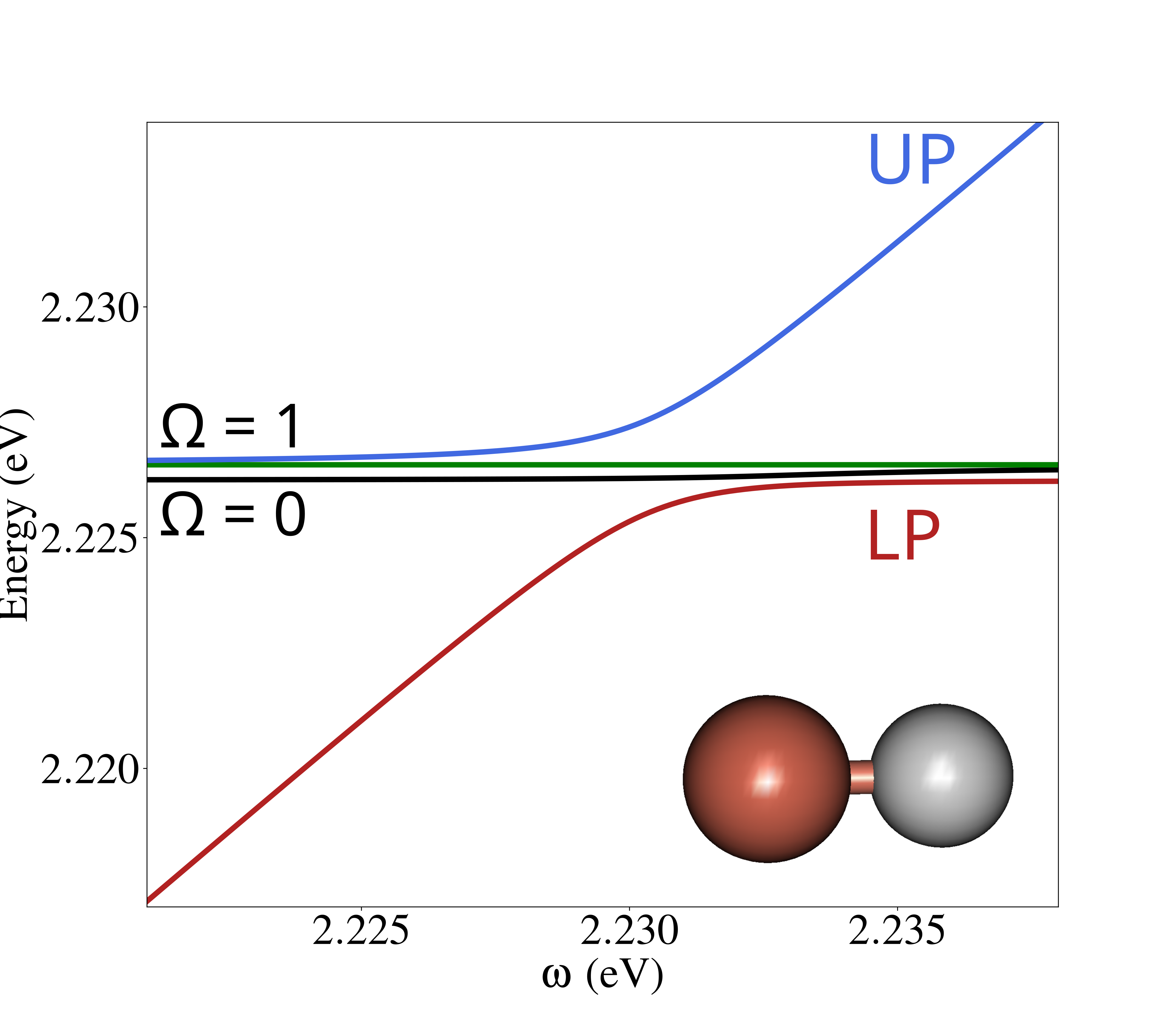}\hfill
    \caption{Excitation energies evaluated with linear response Pol-DHF as a function of the cavity frequency for CuH.}
    \label{fig:spec_CuH}
\end{figure}
In Fig.SI~1, we report the dispersion of the CuH excitation energies with respect to the cavity frequency. Also in this case, despite significantly smaller than the one observed for AuH in Fig. 6 in the main text, we detect an observable Rabi-splitting. This is expected since the Cu atom is much lighter than gold and relativistic effects, including the spin-orbit coupling, are significantly smaller. In this case, the energy difference between the $\Omega= 0$ and $\Omega= 1$ state is 20 times smaller than for AuH ($\sim$ 0.002 eV).
\begin{figure}[h]
    \centering
    \includegraphics[width=1.0\linewidth]{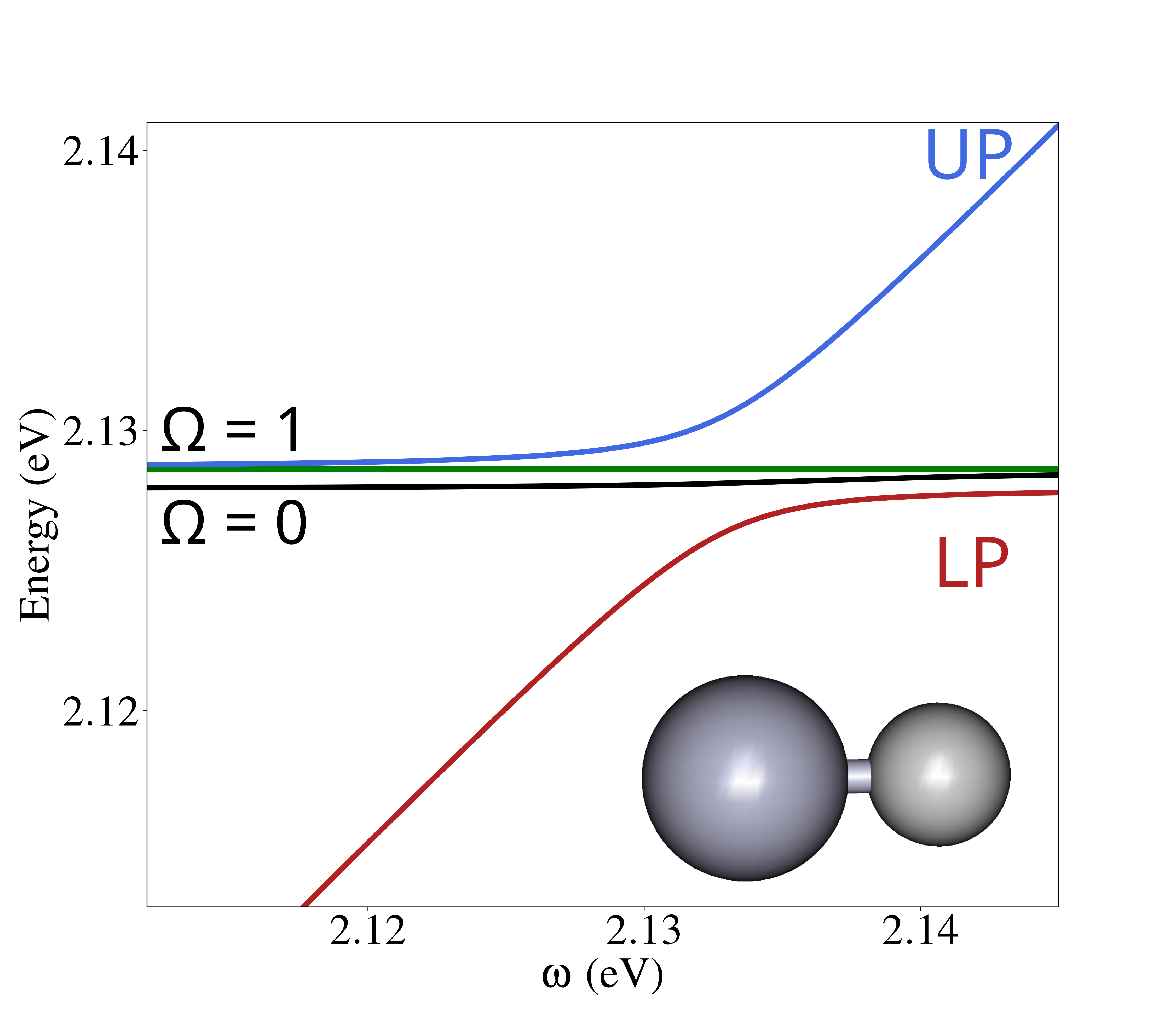}\hfill
    \caption{Excitation energies evaluated with linear response Pol-DHF as a function of the cavity frequency for AgH.}
    \label{fig:spec_AgH}
\end{figure}
\smallbreak
Similar results can be observed also in Fig.SI~2 for AgH. In this case, the observed behavior is somehow intermediate between the one of CuH and the one of AuH, consistently with the fact that Ag is heavier than Cu but lighter than Au. For this system the degeneracy between the three states is slightly lifted due to the 4-component treatment. However, contrary to AuH the energy difference between the $\Omega= 0$ and $\Omega= 1$ state is significantly smaller ($\sim$ 0.009 eV). Notice that for both CuH and AgH, the Rabi-splitting is larger than such energy difference.
\section{Effect of using an uncontracted basis set}
As mentioned in the manuscript, in this work, for computational reasons we have used a contracted basis set, though it is not recommended when using a 4-component method. We have performed calculations with an uncontracted basis set on the smallest hydride CuH in order to assess the impact on our results. While the ground state properties have not been significantly affected by this change, the excited states properties were slightly more affected. Indeed, using an uncontracted basis set tends to amplify the effect of the spin-orbit coupling. Therefore, for a system with a low spin-orbit coupling, the Rabi-Splitting while be smaller since it depends on the spin-orbit coupling. However, it is important to stress that the qualitative behavior remains the same and the effect of the spin-orbit coupling and the Rabi-Splitting is comparable for systems like CuH where relativistic effects are expected to be less important.
\begin{table}[ht!]
    \centering
    \begin{tabular}{|c|c|c|c|}
        \hline
          molecule & DHF (Hartree) & Pol-DHF (Hartree) & $\Delta$E$_{\text{Pol}}$ (eV) \\
         \hline
          CuH &  -1653.7276833807055 & -1653.7333641604048   &  -0.1546 \\
         \hline
    \end{tabular}
    \caption{Ground state total energies calculated at the DHF and Pol-DHF level for CuH with an uncontracted basis set. $\Delta$E$_{\text{Pol}} = \text{E}_{\text{Pol-DHF}} - \text{E}_{\text{DHF}} $.  
    }
    \label{tab:DHF_GS_energies}
\end{table}
\begin{table}[ht!]
    \centering
    \begin{tabular}{|c|c|c|c|c|c|c|}
        \hline
          molecule & \multicolumn{2}{c|}{No Pol} & \multicolumn{2}{c|}{Pol}  \\\cline{2-5}
          & Gaunt (eV) & Breit (eV) & Gaunt (eV) & Breit (eV) \\
         \hline
          CuH &  20.2548&  -1.83556 & 20.2547 & -1.83555 \\
         \hline
    \end{tabular}
    \caption{Energy differences between DHF and DHF-Gaunt(-Breit) level for CuH with an uncontracted basis set.
    }
    \label{tab:Tot_energy_RetVSPol}
\end{table}
\begin{figure*}[ht!]
    \centering
    \includegraphics[width=0.8\linewidth]{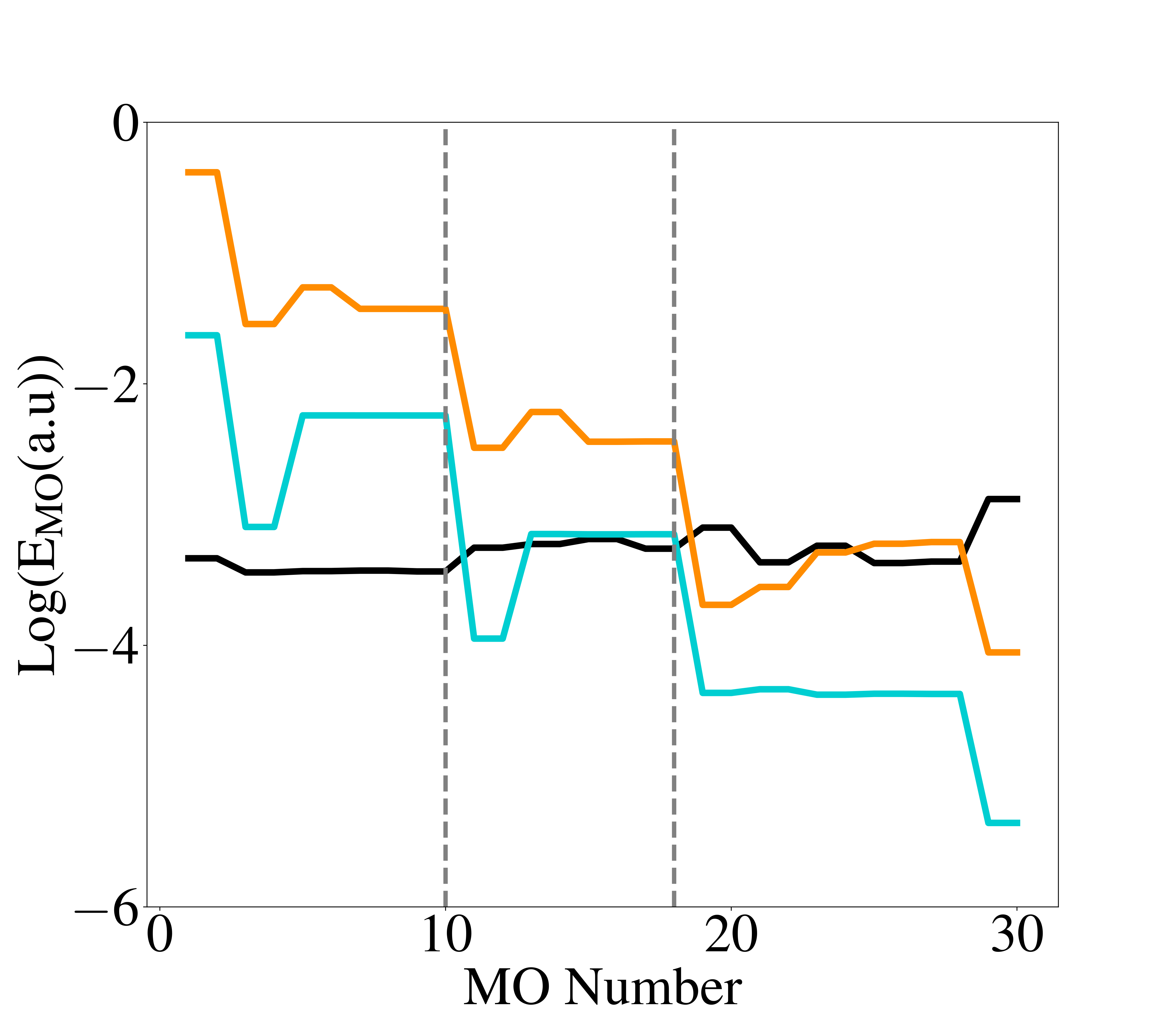}\
    \caption{Absolute contribution of Polaritonic (Black), Gaunt (Orange) and Breit (Cyan) terms on the various molecular orbitals of CuH with and uncontracted basis set. Three zones have been represented on the graph, the one on the left corresponds to the core region, the middle one is an area where Gaunt, Breit and Polaritonic contributions are comparable, and the area on the right to the valence region.}
    \label{fig:orb_mo}
\end{figure*}

\begin{figure}[h]
    \centering
    \includegraphics[width=1.0\linewidth]{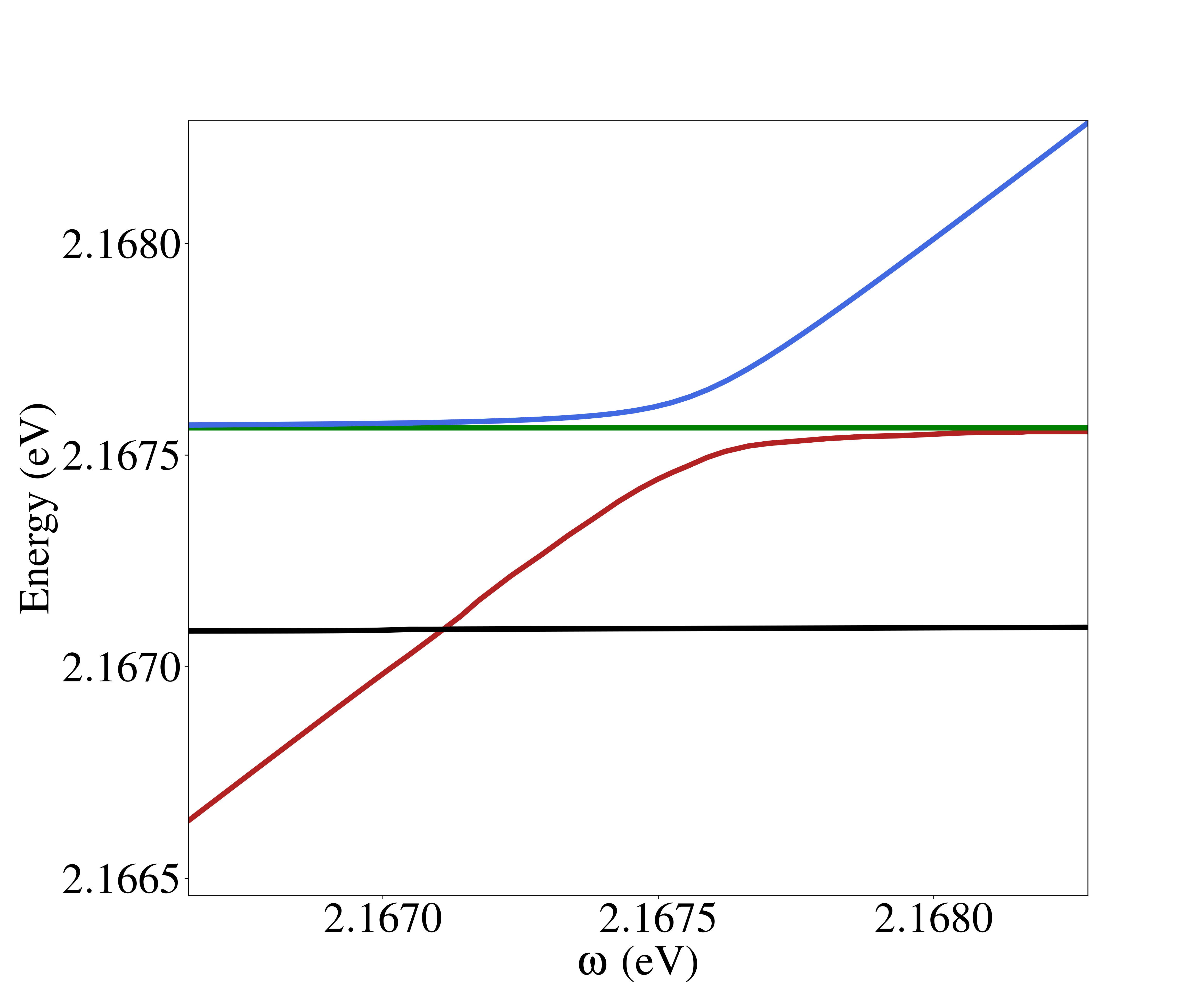}\hfill
    \caption{Excitation energies evaluated with linear response Pol-DHF as a function of the cavity frequency for CuH with an uncontracted basis set.}
    \label{fig:spec_CuH_unc}
\end{figure}




\providecommand{\noopsort}[1]{}\providecommand{\singleletter}[1]{#1}%

\end{document}